%% file: OTH-CPC.tex
\documentclass[preprint,12pt]{elsarticle}

\usepackage{amssymb}

\usepackage{amsmath}
\usepackage{multirow}
\usepackage{fancyvrb}

\newcounter{bla}

\journal{Computer Physics Communications}

\newcommand{\OTH}{{\sc OpTHyLiC}}
\newcommand{\latex}{{\sc LaTeX}}
\newcommand{\pdf}{p.~d.~f.}

\newcommand{\mclimit}{{\sc McLimit}}
\newcommand{\roostats}{{\sc RooStats}}
\newcommand{\histfactory}{{\sc HistFactory}}
\newcommand{\rootcern}{{\sc ROOT}}
\newcommand{\CLs}{\ensuremath{CL_s}}
\newcommand{\CLsb}{\ensuremath{CL_{s+b}}}
\newcommand{\CLb}{\ensuremath{CL_b}}
\newcommand{\mup}{\ensuremath{\mu_{\text{up}}}}
\newcommand{\nochan}{\ensuremath{n}}
\newcommand{\scc}{\ensuremath{s_c}}
\newcommand{\scck}{\ensuremath{s_{c\theta}}}
\newcommand{\bc}{\ensuremath{b_{c}}}
\newcommand{\bck}{\ensuremath{b_{c\theta}}}
\newcommand{\bci}{\ensuremath{b_{ci}}}
\newcommand{\bcik}{\ensuremath{b_{ci\theta}}}
\newcommand{\scnom}{\ensuremath{\scc^{\text{nom}}}}
\newcommand{\scknom}{\ensuremath{\scck^{\text{nom}}}}

\newcommand{\bciknom}{\ensuremath{\bcik^{\text{nom}}}}
\newcommand{\bcnom}{\ensuremath{\bc^{\text{nom}}}}
\newcommand{\bcknom}{\ensuremath{\bck^{\text{nom}}}}

\newcommand{\nobsck}{\ensuremath{\nobs_{c\theta}}}
\newcommand{\snom}{\ensuremath{s^{\text{nom}}}}
\newcommand{\bnom}{\ensuremath{b^{\text{nom}}}}

\newcommand{\nobs}{\ensuremath{N^{\text{obs}}}}
\newcommand{\n}{\ensuremath{N}}
\newcommand{\nc}{\ensuremath{N_c}}
\newcommand{\nck}{\ensuremath{N_{c\theta}}}
\newcommand{\fsyst}{\ensuremath{h^{\text{syst}}}}
\newcommand{\fup}{\ensuremath{h^{\uparrow}}}
\newcommand{\fdown}{\ensuremath{h^{\downarrow}}}
\newcommand{\lhood}{\ensuremath{{\cal L}}}
\newcommand{\lhoodm}{\ensuremath{{\cal L}_\text{m}}}
\newcommand{\dd}{\ensuremath{{\rm d}}}
\newcommand{\dr}{\partial}
\newcommand{\pval}{\text{p-value}}
\newcommand{\qmu}{\ensuremath{q_\mu}}
\newcommand{\qmuobs}{\ensuremath{\qmu^{\text{obs}}}}
\newcommand{\E}[1]{\ensuremath{\mathbb{E}\left[#1\right]}}
\newcommand{\var}[1]{\ensuremath{{\rm var}\left[#1\right]}}

\usepackage{hyperref}

\begin{document}

\begin{frontmatter}

\title{\OTH: an Optimised Tool for Hybrid Limits Computation}

\author[a]{Emmanuel Busato}
\author[a]{David Calvet\corref{author}}
\author[a,b,c]{Timoth\'ee Theveneaux-Pelzer}

\cortext[author] {Corresponding author (\textit{e-mail address:} calvet@in2p3.fr)}
\address[a]{Universit\'e Clermont Auvergne, CNRS/IN2P3, Laboratoire de Physique de Clermont-Ferrand, France}
\address[b]{CPPM, Aix-Marseille Universit\'e, CNRS/IN2P3, France}
\address[c]{now at DESY, Hamburg and Zeuthen, Germany}

\begin{abstract}
A software tool, computing observed and expected upper limits on Poissonian process rates using a hybrid frequentist-Bayesian \CLs~method, is presented.
This tool can be used for simple counting experiments where only signal, background and observed yields are provided or for multi-bin experiments where binned distributions of discriminating variables are provided. 
It allows the combination of several channels and takes into account statistical and systematic uncertainties,
as well as correlations of systematic uncertainties between channels.
It has been validated against other software tools and analytical calculations, for several realistic cases.
\end{abstract}


\end{frontmatter}

\input{Introduction}
\input{Method}

\input{SoftwareDescription}
\input{SoftwareValidation}
\input{Conclusion}
\input{Acknowledgements}

\phantomsection 
\section*{References}
\addcontentsline{toc}{part}{References} 
\bibliographystyle{elsarticle-num}
\bibliography{Biblio}

\end{document}

%% file: Introduction.tex
\section{Introduction}

In a physics experiment, the result must sometimes be interpreted in terms of upper limits on the rate of a particular physical process of interest, dubbed signal.
\OTH\footnote{Pronunciation: [optilik]}~offers an easy-to-use hybrid frequentist-Bayesian solution to set upper limits
for an arbitrary number of channels and backgrounds, using the \CLs~method. 
It can be used for simple counting experiments where only signal, background and observed yields are provided as well as for multi-bin experiments where binned distributions of discriminating variables are provided. 
Statistical and systematic uncertainties are taken into account in a Bayesian way, and correlations
of systematic uncertainties across backgrounds and channels are properly accounted for. \OTH{} can be downloaded from \cite{othweb}.

Other tools such as \mclimit~\cite{mclimit} or \roostats~\cite{roostats} provide the ability to compute limits using the hybrid method.
However, \OTH~was specifically optimized for this method:
it is lightweight, simple and
faster than the tools most often used currently (even in complex cases with many backgrounds, channels and nuisance parameters, limits are typically computed in one minute or less).
Furthermore, it provides several options to configure the treatment of statistical and systematic uncertainties.
\OTH~does not support profiling of uncertainties;
the motivations for this choice are discussed in the next section.

The document is organized as follows.
Typical situations in which \OTH~can be used are described in Sec.~\ref{sec:Purpose}.
The notations used in this paper are introduced in Sec.~\ref{sec:Notations}.
The statistical method implemented in \OTH~is described in Sec.~\ref{sec:MethodDescr}.
The software is described in Sec.~\ref{sec:SoftDescr} and its validation in Sec.~\ref{sec:SoftValid}.

\section{Purpose of the software}
\label{sec:Purpose}

The \OTH~software was written in order to provide users with a lightweight, simple and fast tool to compute upper limits. 
It should be seen as a complementary tool to the several existing ones. 
Because of the absence of profiling, it does not provide the best upper limits in cases where data capable of strongly constraining nuisance parameters exist. 
In such cases, users should prefer other tools like pure frequentist ones (which implement profiling of uncertainties) or pure Bayesian ones. 
\OTH~is more specifically dedicated to situations where no such data exist or to situations where pure frequentist or pure Bayesian tools are too slow for the users' needs. 
In the former situation, it is often observed that pure frequentist, pure Bayesian and hybrid approaches yield similar limits. 
It can therefore be advantageous to use \OTH~rather than tools that are more complicated to handle and slower. 
The latter situation occurs for example when the expected number of events is small and asymptotic properties of the profile likelihood ratio can't be used to speed up pure frequentist calculations (in this case pseudo-experiments must be drawn in order to perform the hypothesis tests which can be very time consuming when profiling a large number uncertainties).
It can also occur when the aim is rather to perform an optimisation of the analysis (to determine the best event selection criteria or the best histogram binning for instance) than to compute final limits. 
Such optimisation studies often do not require high accuracy and can be extremely time consuming or even impossible to do with profiling. 

\OTH~can be used by physicists needing a tool relatively easy to master that provides reliable results taking into account several analysis channels and/or systematic uncertainties. 
It can for example be useful to perform phenomenological sensitivity studies or to recast published searches with newer theoretical models.

\OTH~can also be used by beginners in statistics willing to perform limit calculations in realistic cases for the first time without having to understand first all the details and subtleties of profiling in pure frequentist calculations. 
Those aiming at understanding the details and subtleties of pure frequentist calculations can also use \OTH~as an intermediate step towards that final objective. 
Many of the quantities, definitions and concepts presented in this paper and used in \OTH~are indeed also used by tools like \histfactory~and \roostats~to perform pure frequentist calculations. 
\enlargethispage{1cm}
Even though \OTH~was originally designed for high energy physics it can also be used in other fields such as nuclear measurements or medical physics. 
Many measurements performed in such fields are poissonian and can be modelled with the \OTH~statistical model. 
\OTH~can therefore be used to perform statistical inference and gain valuable informations about physics parameters relevant in those fields too.

\section{Notations and definitions}
\label{sec:Notations}

The following notations will be used throughout the document:
\begin{itemize}
\item $\mu$: signal strength modifier, defined as the tested signal rate divided by a fixed reference signal rate value
(such as obtained from a theoretical prediction);
\item $\mup$: upper limit on the signal strength modifier;
\item $\nochan$: number of channels;
\item $\scck$: event yield for signal process in channel $c$ ($c\in[1,\nochan]$) and bin $\theta$;
\item $\scknom$: nominal reference event yield for signal process in channel $c$ and bin $\theta$;
\item $\sigma_{cl}$: absolute systematic uncertainty for signal process in channel $c$ and bin $\theta$, due to the finite size of the control sample;
\item $\bcik$: event yield for type $i$ background process in channel $c$ and bin $\theta$;
\item $\bciknom$: nominal event yield for type $i$ background process in channel $c$ and bin $\theta$;
\item $\bcknom=\sum\limits_{i\in\mathrm{backgrounds}}\bciknom$: total nominal yield for background processes in channel $c$ and bin $\theta$;
\item $\sigma_{ci\theta}$: absolute systematic uncertainty for type $i$ background process in channel $c$ and bin $\theta$, due to the finite size of the control sample;
\item $\nck$: event yield in channel $c$ and bin $\theta$;
\item $\nobsck$: event yield actually observed in the data or the pseudo-data in channel $c$ and bin $\theta$.
\end{itemize}

A channel corresponds to a region enriched in signal events (also called signal region). 
In the case of multi-bin experiments, the user must provide binned distributions (or histograms) for each channel, background and signal. 
Bins of these distributions are referred to by the index $\theta$.
For counting experiments, only yields for each background and signal in each channel are needed.
In this case, the index $\theta$ is irrelevant and is discarded.

%% file: Method.tex
\section{Method description}
\label{sec:MethodDescr}

\OTH~implements the \CLs~method~\cite{0954-3899-28-10-313}.
Pseudo-experiments are generated and the distribution of a test statistic is determined under both signal plus background and background only hypotheses.
From these distributions, \CLs~is computed and the upper limit on the signal strength modifier \mup~is found by solving the equation
\begin{equation}
\label{eq:muUpFromCLs}
\CLs\left(\mu\right)=\alpha,
\end{equation}
for a fixed value of $\alpha$ which is set depending on the chosen confidence level $1-\alpha$.
For 95\% confidence level upper limits, $\alpha=0.05$.
The test statistic used is the ratio of likelihoods under signal plus background and background only hypotheses:
\begin{equation}
\label{eq:TestStat}
\qmu=-2\ln\frac{{\lhood}(\mu)}{\lhood(\mu=0)}.
\end{equation}

The general form of the likelihood is discussed in Sec.~\ref{sec:StatModel}.
The statistical and systematic uncertainties are included by the use of nuisance parameters, as detailed
in Sec.~\ref{sec:StatUncertTreatment} and~\ref{sec:SystUncertTreatment}, respectively.
These uncertainties are accounted for in a Bayesian way: the inference of the observed upper limit is performed
from pseudo-experiments using the distribution of the likelihood marginalised over the nuisance parameters, as explained in Sec.~\ref{sec:InferenceMuup}.
The procedure for calculating the expected limits is summarised in Sec.~\ref{sec:ExpectedLimits}.

\subsection{Statistical model}
\label{sec:StatModel}

The full likelihood, including all nuisance parameters, is given by:
\begin{equation}
\label{eq:fullLhood}
\begin{array}{rl}
\lhood(\mu,\{\scck',\bcik',\eta_j\})=\prod\limits_{c,\theta}&\left[\frac{\left(\mu\scck+\bck\right)^{\nck}}{\nck!}e^{-\left(\mu\scck+\bck\right)}\right.\\
&\times f\left(\scck';\scknom,\sigma_{c\theta}\right)\prod\limits_{i}f\left(\bcik';\bciknom,\sigma_{ci\theta}\right)\left.\vphantom{\frac{\left(\mu\scck+\bck\right)^{\nck}}{\nck!}}\right]\prod\limits_{j}g\left(\eta_j\right),\\
\end{array}
\end{equation}
where:
\begin{itemize}
\item the index $c$ runs over the channels;
\item the index $i$ runs over the backgrounds;
\item the index $\theta$ runs over the bins;
\item the index $j$ runs over the systematic uncertainties;
\item $\scck=\scck'\times k_{c\theta}^{\text{syst}}\left(\{\eta_j\}\right)$;
\item $\bck=\sum\limits_{i\in\text{backgrounds}}\bcik=\sum\limits_{i\in\text{backgrounds}}\bcik'\times k_{ci\theta}^{\text{syst}}\left(\{\eta_j\}\right)$.
\end{itemize}

In \OTH{}, the signal always adds to the background ($\mu\scck$ is always positive). Cases where the searched signal causes a deficit of events with respect to the background only hypothesis therefore can't be treated with \OTH{}.

The set of nuisance parameters $\{\scck',\bcik',\eta_j\}$ can be divided into two categories.
The first ones, $\{\scck',\bcik'\}$, account for the systematic uncertainties due to the finite size of the control samples
used to estimate the signal and background nominal yields \scknom~and \bciknom, respectively.
Such uncertainties are referred to as ``statistical uncertainties'' in this paper. These nuisance parameters are constrained by the functions $f$.
The second ones, $\{\eta_j\}$, account for the systematic uncertainties. They are constrained by the functions $g$.
The variation of the signal and background yields under the effect of the systematic uncertainties
are described by the functions $k_{c\theta}^{\text{syst}}\left(\{\eta_j\}\right)$ and $k_{ci\theta}^{\text{syst}}\left(\{\eta_j\}\right)$, respectively.

Bins of discriminating variable histograms are treated as channels: the drawing of Poisson
random numbers and the statistical uncertainties on the yield of each process in each bin are independent,
but a given systematic uncertainty may affect the yields in a correlated way.
It is therefore equivalent to run a counting experiment with $N$ channels and a binned experiment with $N$ bins where the yields and uncertainties in each bin match those in the channels.

\subsection{Treatment of statistical uncertainties}
\label{sec:StatUncertTreatment}

The nuisance parameters $\scc'$ and $\bci'$, accounting for the statistical uncertainties on the nominal signal and background yields,
are constrained by probability density functions (\pdf) $f$ of the form:
\begin{equation}
\label{eq:statConstraintPdfGeneralForm}
f(y;y^{\text{nom}},\sigma),
\end{equation}
where $y$ is the nuisance parameter, $y^{\text{nom}}$ the nominal yield and $\sigma$
the statistical uncertainty, which can be the square root of summed squared weights
(e.g. when $y^{\text{nom}}$ is estimated from a mixture of normalised simulated samples).

In \OTH, five different \pdf\ can be used as constraints:
a normal distribution, a log-normal distribution, or three different types of gamma distributions.

\subsubsection{Normal and log-normal constraints}
\label{sec:NormalLogNormal}

The parameters of the normal and log-normal \pdf\ are chosen so that the average and the standard deviation
of these distributions are equal to $y^{\text{nom}}$ and $\sigma$, respectively.
The normal distribution has the form:
\begin{equation}
\label{eq:normalDistribGeneralForm}
f_\text{N}(y;y^{\text{nom}},\sigma)=\frac{1}{\sigma\sqrt{2\pi}}\exp\left(-\frac{(y-y^{\text{nom}})^2}{2\sigma^2}\right),
\end{equation}
and the log-normal distribution has the form:
\begin{equation}
\label{eq:lognormalDistribGeneralForm}
f_\text{L}(y;y^{\text{nom}},\sigma)=\frac{1}{y\ b\sqrt{2\pi}}\exp\left(-\frac{(\ln y-a)^2}{2 b^2}\right),
\end{equation}
with:
\begin{equation}
\label{eq:lognormalDistribParamDef}
a=\ln\hspace{-2.5pt}\left(\hspace{-2.5pt}\frac{\left(y^{\text{nom}}\right)^2}{\sqrt{\left(y^{\text{nom}}\right)^2+\sigma^2}}\hspace{-2.5pt}\right)\text{and}\
b=\sqrt{\hspace{-2.5pt}\ln\hspace{-2.5pt}\left(\hspace{-2.5pt}1+\frac{\sigma^2}{\left(y^{\text{nom}}\right)^2}\hspace{-2.5pt}\right)}.
\end{equation}
The normal distribution can be defined for any real $y$ values, including non-physical negative event yields.
On the contrary, the log-normal distribution is defined only for \mbox{$y>0$.}

In the log-normal case, the nominal yield~$y^{\text{nom}}$ and statistical uncertainty~$\sigma$ estimated by users are interpreted as the average and standard deviation of the distribution, respectively.
However, a different interpretation is possible.
For example, if the statistical uncertainty results from a maximum likelihood estimate on some auxiliary dataset, $y^{\text{nom}}$ corresponds to the mode of the likelihood and $\sigma$ to the uncertainty as measured from the curvature of the log-likelihood around the mode ($\sigma = -1/\sqrt{\left(\ln f_\text{L}\right)^{''}\left(y=y^{\text{nom}}\right)}$).
The parameters $a$ and $b$ of the log-normal distribution should therefore be equal to:
\begin{equation}
\label{Eq:AltLogNaAndbParameters}
a = \frac{\sigma^2}{\left(y^{\text{nom}}\right)^2} + \ln\left(y^{\text{nom}}\right)\text{and}\
b = \frac{\sigma}{y^{\text{nom}}}.
\end{equation}
This can be achieved in \OTH\ by setting input parameters to:
\begin{equation}
e^{a+\frac{b^2}{2}}~\text{and}\
e^{a+\frac{b^2}{2}}\sqrt{e^{b^2} - 1}~\text{,}
\end{equation}
with $a$ and $b$ given by Eq.~\ref{Eq:AltLogNaAndbParameters}, rather than to $y^{\text{nom}}$ and $\sigma$, respectively.

\subsubsection{Gamma constraints}
\label{sec:Gamma}

Three gamma \pdf\ are implemented in \OTH. They have the general form:
\begin{equation}
\label{eq:gammaDistribGeneralForm}
f_\text{G}(y;a,b)=\frac{a\left(ay\right)^{b-1}e^{-ay}}{\Gamma\left(b\right)},
\end{equation}
where $a$ and $b$ are the rate and shape parameters respectively.
As for the log-normal, the gamma distributions are defined for any strictly positive value of $y$.

This \pdf\ can be seen as the posterior distribution obtained from an auxiliary measurement
accounting for the Poissonian nature of the statistical uncertainty.
Accounting for this nature is not straightforward since events are in general weighted.
A popular approach (used for example in \histfactory~\cite{Cranmer:1456844}) is to consider an imaginary auxiliary measurement
in which all events have unit weight (which can therefore be described by a Poisson distribution) 
and in which the relative statistical uncertainty is equal to that used in the main measurement ($\sigma/y^{\text{nom}}$).

Let $N_\text{aux}$ be the number of events observed in this auxiliary measurement.
The auxiliary measurement likelihood is:
\begin{equation}
P(N_\text{aux};\lambda)=\frac{\lambda^{N_\text{aux}}}{\Gamma\left(N_\text{aux}+1\right)}e^{-\lambda},
\end{equation}
where $\lambda$ is the (unknown) nuisance parameter.
It can be written as:
\begin{equation}
\lambda=\gamma \ N_\text{aux}^{\text{nom}},
\end{equation}
where $N_\text{aux}^{\text{nom}}$ is the nominal value of $N_\text{aux}$ and $\gamma$ the nuisance parameter
affecting the yield in the main measurement -- the product of Poisson terms in Eq.~\ref{eq:fullLhood} -- in a multiplicative way: $y=\gamma \ y^{\text{nom}}$.
$N_\text{aux}^\text{nom}$ is found by imposing that the relative statistical uncertainty
is the same in the auxiliary and main measurements:
\begin{equation}
N_\text{aux}^{\text{nom}}=\left(\frac{y^{\text{nom}}}{\sigma}\right)^2.
\end{equation}
The auxiliary measurement likelihood can then be written as follows:
\begin{equation}
\label{eq:auxiliaryLikelihoodWithGamma}
P(N_\text{aux};\gamma)=\frac{\left(\gamma\left(y^{\text{nom}}/\sigma\right)^2\right)^{N_\text{aux}}}{\Gamma\left(N_\text{aux}+1\right)}e^{-\gamma \left(y^{\text{nom}}/\sigma\right)^2}.
\end{equation}
From this likelihood, posterior distributions for $\gamma$ can be determined for various prior distributions $\pi\left(\gamma\right)$:
\begin{equation}
g\left(\gamma\right)=\frac{P(N_\text{aux}=N_\text{aux}^{\text{nom}};\gamma)\pi\left(\gamma\right)}{\displaystyle\int P(N_\text{aux}=N_\text{aux}^{\text{nom}};\gamma)\pi\left(\gamma\right)\dd\gamma}.
\end{equation}
These posterior distributions can then be translated into posterior distributions for the yield $y$
(hereafter denoted as $f(y;y^{\text{nom}},\sigma)$, using the same notation as in Eq.~\ref{eq:statConstraintPdfGeneralForm}).

Three prior distributions are considered; in each case, the posterior distribution for $y$ is a gamma distribution
with the general form given by Eq.~\ref{eq:gammaDistribGeneralForm}:
\begin{subequations}
\begin{itemize}
\item $\pi\left(\gamma\right)\propto 1$ (uniform prior); in this case, the posterior distribution is:
\begin{equation}
\label{eq:gammaPosteriorUni}
f(y;y^{\text{nom}},\sigma)=f_\text{G}\left(y;a=y^{\text{nom}}/\sigma^2,b=\left(y^{\text{nom}}/\sigma\right)^2+1\right);
\end{equation}
\item $\pi\left(\gamma\right)\propto 1/\sqrt{\gamma}$ (Jeffreys prior\footnote{We recall that Jeffreys prior
is $\pi\left(\gamma\right)\propto\sqrt{I(\gamma)}$, where the Fisher information is $I(\gamma)=\E{\left(\frac{\dr\log P\left(N_\text{aux};\gamma\right)}{\dr\gamma}\right)^2}$.
Injecting Eq.~\ref{eq:auxiliaryLikelihoodWithGamma} in this expression leads to $\pi\left(\gamma\right)\propto 1/\sqrt{\gamma}$.}); in this case, the posterior distribution is:
\begin{equation}
\label{eq:gammaPosteriorJeffrey}
f(y;y^{\text{nom}},\sigma)=f_\text{G}\left(y;a=y^{\text{nom}}/\sigma^2,b=\left(y^{\text{nom}}/\sigma\right)^2+1/2\right);
\end{equation}
\item $\pi\left(\gamma\right)\propto 1/\gamma$ (hyperbolic prior); in this case, the posterior distribution is:
\begin{equation}
\label{eq:gammaPosteriorInv}
f(y;y^{\text{nom}},\sigma)=f_\text{G}\left(y;a=y^{\text{nom}}/\sigma^2,b=\left(y^{\text{nom}}/\sigma\right)^2\right).
\end{equation}
\end{itemize}
\end{subequations}

The three gamma constraints available in \OTH~correspond to the three posteriors given
in Eq.~\ref{eq:gammaPosteriorUni}, \ref{eq:gammaPosteriorJeffrey} and~\ref{eq:gammaPosteriorInv}.
Their parameters are summarized in Tab.~\ref{tab:gammaConstraints}.
\begin{table*}[!htb]\centering\footnotesize
\renewcommand{\arraystretch}{1.3}
\begin{tabular}{ccc}
\hline
\multirow{2}{*}{prior}    & \multicolumn{2}{c}{posterior parameters} \\ 
 & $a$   &    $b$   \\ \hline
$\pi\left(\gamma\right)\propto 1$  & $y^{\text{nom}}/\sigma^2$  & $\left(y^{\text{nom}}/\sigma\right)^2+1$ \\ \hline
$\pi\left(\gamma\right)\propto 1/\sqrt{\gamma}$  & $y^{\text{nom}}/\sigma^2$  & $\left(y^{\text{nom}}/\sigma\right)^2+1/2$ \\ \hline
$\pi\left(\gamma\right)\propto 1/\gamma$  & $y^{\text{nom}}/\sigma^2$  & $\left(y^{\text{nom}}/\sigma\right)^2$ \\ \hline
\end{tabular}
\caption{Summary of the three gamma constraints available in \OTH. The constants $a$ and $b$ are parameters of the posterior distribution whose general form is given in Eq.~\ref{eq:gammaDistribGeneralForm}.\label{tab:gammaConstraints}}
\end{table*}

\subsubsection{Comparison of the constraint functions}
\label{sec:ConstraintChoice}

\begin{figure*}[!htb]
\begin{center}
\hspace*{-0.6cm}
\includegraphics[width=\textwidth]{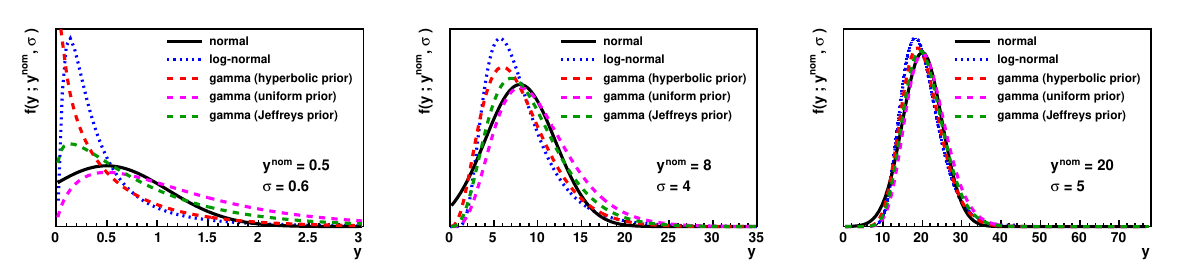}
\caption{Comparison of normal, log-normal and gamma probability density functions used for statistical uncertainties for three values of $y^{\text{nom}}$ and~$\sigma$.\label{fig:plotNormalLogNGamma}}
\end{center}
\end{figure*}

The five available constraint functions -- normal, log-normal and the three gamma distributions -- are compared
for three different values of $y^{\text{nom}}$ and $\sigma$ in Fig.~\ref{fig:plotNormalLogNGamma}.
When $\sigma$ is small with respect to $y^{\text{nom}}$, the distributions are very close to each other.
Otherwise, the differences between the distributions can be significant.

Table~\ref{tab:PDFComparisons} compares the mode, the expectation and standard deviation for the five available constraint functions.
For the normal and log-normal distribution, the expectation and standard deviation are equal to the nominal yield and statistical uncertainty (see Sec.~\ref{sec:NormalLogNormal}).
For the gamma distributions the expectation and standard deviation are equal to the nominal yield and statistical uncertainty
only for the hyperbolic prior $\pi\left(\gamma\right)\propto 1/\gamma$.
For the gamma constraints with the uniform and Jeffreys priors, it is not the case, but the difference vanishes in the asymptotic limit.
\begin{table*}[!htb]\centering\footnotesize
\renewcommand{\arraystretch}{1.3}
\begin{tabular}{cccc}
\hline
distribution    & mode  &  $\E{y}$  &  $\var{y}$\\ \hline
normal & $y^{\text{nom}}$ & $y^{\text{nom}}$ & $\sigma^2$ \\ \hline
log-normal & $y^{\text{nom}}\left(1+\left(\frac{\sigma}{y^{\text{nom}}}\right)^2\right)^{-3/2}$ & $y^{\text{nom}}$ & $\sigma^2$ \\ \hline
gamma  & \multirow{2}{*}{$y^{\text{nom}}$}  &  \multirow{2}{*}{$y^{\text{nom}}\left(1+\left(\frac{\sigma}{y^{\text{nom}}}\right)^2\right)$}  &  \multirow{2}{*}{$\sigma^2\left(1+\left(\frac{\sigma}{y^{\text{nom}}}\right)^2\right)$}   \\ 
(uniform prior) & & & \\ \hline
gamma  & \multirow{2}{*}{$y^{\text{nom}}\left(1-\frac{1}{2}\left(\frac{\sigma}{y^{\text{nom}}}\right)^2\right)$}  & \multirow{2}{*}{$y^{\text{nom}}\left(1+\frac{1}{2}\left(\frac{\sigma}{y^{\text{nom}}}\right)^2\right)$} & \multirow{2}{*}{$\sigma^2\left(1+\frac{1}{2}\left(\frac{\sigma}{y^{\text{nom}}}\right)^2\right)$} \\ 
(Jeffreys prior) & & & \\ \hline
gamma & \multirow{2}{*}{$y^{\text{nom}}\left(1-\left(\frac{\sigma}{y^{\text{nom}}}\right)^2\right)$}  & \multirow{2}{*}{$y^{\text{nom}}$} & \multirow{2}{*}{$\sigma^2$} \\ 
(hyperbolic prior) & & & \\ \hline
\end{tabular}
\caption{Comparison of the mode, expectation, and standard deviation for the five constraints available in \OTH.\label{tab:PDFComparisons}}
\end{table*}

Care should be taken when statistical uncertainties are large compared with the nominal yields or when nominal yields are equal to zero.
When statistical uncertainties are large, constraint \pdf\ can be truncated at zero in the normal case.
In such cases, log-normal and gamma should be preferred.
When nominal yields are equal to zero, log-normal and gamma are undefined.
In such cases, \OTH\ automatically selects a normal constraint truncated at zero.
In general, users are encouraged to study the stability of the limit calculations depending on the assumptions made in such circumstances.

\subsection{Treatment of systematic uncertainties}
\label{sec:SystUncertTreatment}

Systematic uncertainties on nominal yields \scknom\ and \bciknom\ are accounted for by including the set of nuisance parameters $\{\eta_j\}$.
They are assumed to be either 100\% correlated or completely uncorrelated.
The total number of nuisance parameters is therefore equal to the total number of independent systematic uncertainties,
including all channels, backgrounds and signals.
The correlation factor between two nuisance parameters $\eta_j$ and $\eta_k$ is given by the Kronecker symbol $\delta_{jk}$.
The term constraining nuisance parameters in the likelihood can thus be factorized into the product of individual constraint terms.
In Eq.~\ref{eq:fullLhood}, the nuisance parameter for each systematic uncertainty of index $j$ is constrained by a standard normal \pdf\ $g$ of the form:
\begin{equation}
g\left(\eta_j\right)=\frac{1}{\sqrt{2\pi}}e^{-\frac{\eta_j^2}{2}}.
\end{equation}

As stated in Sec.~\ref{sec:StatModel}, the effect of systematic uncertainties on the yield is described by relations of the form:
\begin{equation}
y=y^\text{nom}\times k^{\text{syst}}\left(\{\eta_j\}\right),
\end{equation}
where $y$ is the varied yield, $y^\text{nom}$ the nominal yield and $k^{\text{syst}}\left(\{\eta_j\}\right)$
the function describing the variation of the yield with the set of nuisance parameters $\{\eta_j\}$.
\OTH~provides two different solutions for combining the effect of multiple nuisance parameters:
\begin{itemize}
\item additive:
\begin{subequations} 
\label{eq:systCombiMode}
  \begin{equation}
k^{\text{syst}}\left(\{\eta_j\}\right)-1=\sum\limits_{j\in\text{systematics}}\left[\fsyst_j\left(\eta_j\right)-1\right]; \label{eq:additiveCombi}
  \end{equation}
\item multiplicative: 
  \begin{equation}
k^{\text{syst}}\left(\{\eta_j\}\right)=\prod\limits_{j\in\text{systematics}}\fsyst_j\left(\eta_j\right). \label{eq:multiplicativeCombi}
  \end{equation}
\end{subequations}
\end{itemize}
In Eq.~\ref{eq:additiveCombi} and~\ref{eq:multiplicativeCombi}, $\fsyst_j\left(\eta_j\right)$ is the function
describing the variation of the yield with nuisance parameter $j$.
In both cases, when the effect of only one nuisance parameter is considered, $k^{\text{syst}}\left(\eta_j\right)=\fsyst_j\left(\eta_j\right)$ as it should.

For each systematic uncertainty $j$, the corresponding nuisance parameter $\eta_j$ is chosen such that $\eta_j=0$ corresponds to no variation,
$\eta_j=+1$ to a $+1\sigma$ variation, and $\eta_j=-1$ to a $-1\sigma$ variation.
The main issue associated to systematic uncertainties is the choice of the function $\fsyst_j\left(\eta_j\right)$ that relates the effect
of the systematic uncertainty to its associated nuisance parameter.
Usually, $\fsyst_j$ is known, for each systematic, for $\eta_j=0, -1$ and $+1$.
The value of $h_j^{\mathrm{sys}}$ for $\eta_j=0$ is by definition equal to 1: it corresponds to the case $y=y^\text{nom}$.
Let $\fup_j$ ($\fdown_j$) be the relative variation of the yield when systematic $j$ is varied by $+1$ ($-1$) $\sigma$.
Thus, for all $j$:
\begin{subequations} 
\label{eq:systFuncConstraint}
  \begin{gather}
    \fsyst_j\left(\eta_j=0\right)=1; \label{eq:systFuncConstraint1} \\
    \fup_j=\fsyst_j\left(\eta_j=+1\right)-1; \label{eq:systFuncConstraint2} \\
    \fdown_j=\fsyst_j\left(\eta_j=-1\right)-1. \label{eq:systFuncConstraint3}
  \end{gather}
\end{subequations}

\begin{figure*}[!htb]
\begin{center}
\includegraphics[scale=0.7]{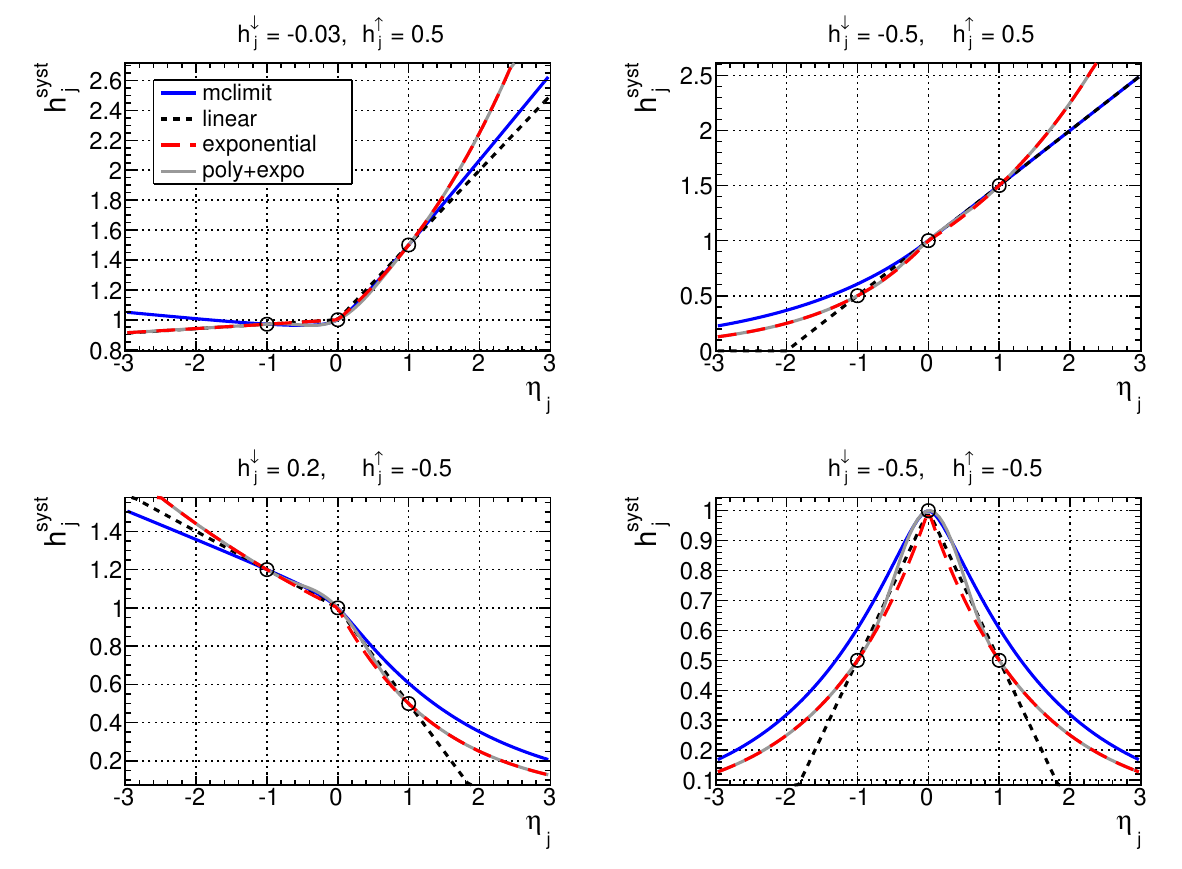}
\caption{Illustration of interpolation and extrapolation functions implemented in \OTH~for $\fdown_j=-0.03$ and $\fup_j=0.5$ (top left),
$\fdown_j=-0.5$ and $\fup_j=0.5$ (top right), $\fdown_j=0.2$ and $\fup_j=-0.5$ (bottom left) and $\fdown_j=-0.5$ and $\fup_j=-0.5$ (bottom right).
\label{fig:ExampleFunctionsInterExtrap}}
\end{center}
\end{figure*}

The problem consists in finding continuous functions $\fsyst_j\left(\eta_j\right)$, interpolating
for $\eta_j\in\left[-1,+1\right]$ and extrapolating for $\eta_j>+1$ and $\eta_j<-1$,
and such that Eqs.~\ref{eq:systFuncConstraint} are satisfied -- at least approximately.
Four choices are currently available in \OTH, and are represented in Fig.~\ref{fig:ExampleFunctionsInterExtrap}:
\begin{itemize}
\item piece-wise linear interpolation and extrapolation, defined in Sec.~\ref{sec:LinearInterExtrap};
\item piece-wise exponential interpolation and extrapolation, defined in Sec.~\ref{sec:ExpoInterExtrap};
\item polynomial interpolation and exponential extrapolation, defined in Sec.~\ref{sec:PolyInterExpoExtrap};
\item ``\mclimit''\ interpolation and extrapolation, defined in Sec.~\ref{sec:McLimitInterExtrap}
and corresponding to the choice followed in the \mclimit\ program.
\end{itemize}

\subsubsection{Piece-wise linear interpolation and extrapolation}
\label{sec:LinearInterExtrap}

The piece-wise linear interpolation and extrapolation is defined as follows:
\begin{equation}
\fsyst_j\left(\eta_j\right)=
\left\{
\begin{matrix}
1+\eta_j\fup_j\quad\text{if $\eta_j$ is positive}; \vspace*{0.15cm}\\
1-\eta_j\fdown_j\quad\text{if $\eta_j$ is negative}.
\end{matrix}
\right.
\end{equation}
From this definition, it appears that $\fsyst_j$ can be negative if $\fup_j$ or $\fdown_j$ is negative.
This unphysical behaviour is dealt with in \OTH\ by setting $\fsyst_j$ to zero when it occurs.

\subsubsection{Piece-wise exponential interpolation and extrapolation}
\label{sec:ExpoInterExtrap}

The piece-wise exponential interpolation and extrapolation is defined as follows:
\begin{equation}
\fsyst_j\left(\eta_j\right)=
\left\{
\begin{matrix}
\left(1+\fup_j\right)^{\eta_j}\quad\text{if $\eta_j$ is positive}, \vspace*{0.15cm}\\
\left(1+\fdown_j\right)^{-\eta_j}\quad\text{if $\eta_j$ is negative}.
\end{matrix}
\right.
\end{equation}
From this definition, it appears that $\fsyst_j$ can be negative if $\fup_j$ or $\fdown_j$ is lower than -1.
This unphysical behaviour is dealt with in \OTH\ by applying a linear interpolation and extrapolation instead
(see Sec. \ref{sec:LinearInterExtrap}) when it occurs.
If both $\fup_j$ and $\fdown_j$ are greater than -1, $\fsyst_j$ is positive for all $\eta_j$ values.

\subsubsection{Polynomial interpolation and exponential extrapolation}
\label{sec:PolyInterExpoExtrap}

The polynomial interpolation and exponential extrapolation is defined as follows:
\begin{equation}
\fsyst_j\left(\eta_j\right)=
\left\{
\begin{matrix}
\left(1+\fup_j\right)^{\eta_j}\quad\text{if $\eta_j\geq 1$}, \vspace*{0.15cm}\\
1+\sum\limits_{i=1}^6a_i\eta_j^i\quad\text{if $-1<\eta_j<1$}, \vspace*{0.15cm}\\
\left(1+\fdown_j\right)^{-\eta_j}\quad\text{if $\eta_j\leq -1$},
\end{matrix}
\right.
\end{equation}
where the coefficients $a_i$ are chosen such that $\fsyst_j\left(\eta_j\right)$ and its first and second derivatives
are continuous at $|\eta_j|=1$ and $\eta_j=0$.

\subsubsection{``\mclimit'' interpolation and extrapolation}
\label{sec:McLimitInterExtrap}

The ``\mclimit'' interpolation and extrapolation is defined as follows:
\begin{equation}
\fsyst_j\left(\eta_j\right)=
\left\{
\begin{matrix}
1+B\quad\text{if $B\geq0$}, \vspace*{0.15cm}\\
e^B\quad\text{if $B<0$},
\end{matrix}
\right.
\end{equation}
where:
\begin{equation}
B=
\left\{
\begin{matrix}
\eta_j\fup_j\left(1-R\right)+RQ\quad\text{if $\eta_j$ is positive}, \vspace*{0.15cm}\\
-\eta_j\fdown_j\left(1-R\right)+RQ\quad\text{if $\eta_j$ is negative},
\end{matrix}
\right.
\end{equation}
\begin{equation}
Q=\eta_j\frac{\fup_j-\fdown_j}{2}+\eta_j^2\frac{\fup_j+\fdown_j}{2}\quad\text{and}\quad R=\frac{1}{1+3|\eta_j|}.
\end{equation}
This definition ensures that the first derivative of $\fsyst_j$ at $\eta_j=0$ is continuous and that $\fsyst_j>0$ for all $\eta_j$ values.
Equation~\ref{eq:systFuncConstraint2} (\ref{eq:systFuncConstraint3}) is exactly satisfied when $\fup_j>0$ ($\fdown_j>0$)
but only to first order when $\fup_j<0$ ($\fdown_j<0$).
Indeed, the following relations hold:
\begin{equation}
\fsyst_j\left(\eta_j=+1\right)=
\left\{
\begin{matrix}
e^{\fup_j}\quad\text{if}\quad\fup_j<0, \vspace*{0.15cm}\\
1+\fup_j\quad\text{if}\quad\fup_j\geq0,
\end{matrix}
\right.
\end{equation}
\begin{equation}
\fsyst_j\left(\eta_j=-1\right)=
\left\{
\begin{matrix}
e^{\fdown_j}\quad\text{if}\quad\fdown_j<0, \vspace*{0.15cm}\\
1+\fdown_j\quad\text{if}\quad\fdown_j\geq0.
\end{matrix}
\right.
\end{equation}
Also, when the effect of the considered uncertainty is symmetric ($\fup_j=-\fdown_j$),
this interpolation/extrapolation scheme is equivalent to the piece-wise linear one (see Sec. \ref{sec:LinearInterExtrap})
for $\eta_j>0$ if $\fup\geq0$ or $\eta_j<0$ if $\fdown\geq0$.

\subsection{Inference of observed upper limit $\mup$}
\label{sec:InferenceMuup}

The observed upper limit on the signal strength \mup\ is derived from the \CLs\ method using $\qmu$ (Eq.~\ref{eq:TestStat}) as test statistic.
$\qmu$ is computed using the nominal likelihood:
\begin{equation}
\lhood(\mu)=
\lhood(\mu,\{\scck'\}\hspace{-2.5pt}=\hspace{-2.5pt}\{\scknom\},\{\bcik'\}\hspace{-2.5pt}=\hspace{-2.5pt}\{\bciknom\},\{\eta_j\}=0).
\end{equation}
It thus reduces to the following simple form:
\begin{equation}
\label{eq:TestStatAllChannels}
\qmu=\sum\limits_{c,\theta} \qmu^{c\theta},
\end{equation}
where $\qmu^{c\theta}$ is the test for channel $c$ and bin $\theta$ given by:
\begin{equation}
\label{eq:TestStatChannelc}
\qmu^{c\theta}=2\left(\mu \scknom-\nck\ln\frac{\mu\scknom+\bcknom}{\bcknom}\right).
\end{equation}

The distributions of $\qmu$ under signal plus background and background only hypotheses (hereafter denoted as $p(\qmu|\mu)$ and $p(\qmu|0)$ respectively) 
are determined by generating pseudo-experiments from the marginal likelihood:
\begin{equation}
\label{eq:LhoodMarginal}
\lhoodm\left(\mu\right)=\displaystyle\int\lhood(\mu,\{\scck',\bcik',\eta_j\})~\prod\limits_j\dd\eta_j\prod\limits_{c,\theta}\dd\scck'\prod\limits_i\dd\bcik'.
\end{equation}

\begin{figure}[!htb]
\begin{center}
\hspace*{-0.6cm}
\includegraphics[scale=0.4]{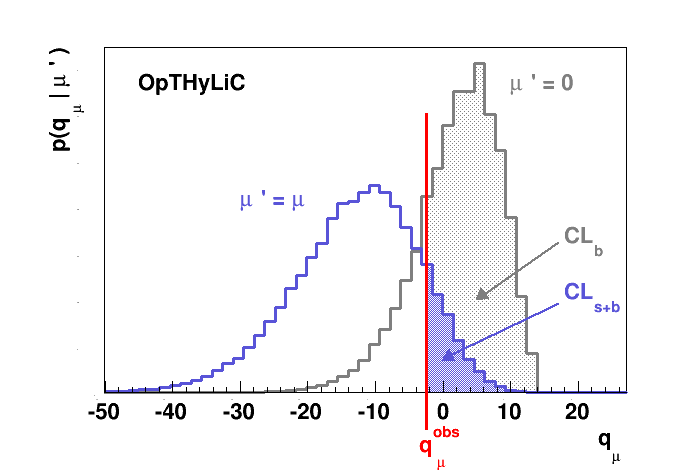}
\caption{Example of distributions of $\qmu$ under signal+background ($\mu'=\mu$) and background only hypotheses ($\mu'=0$).
\label{fig:testStatDistribExample}}
\end{center}
\end{figure}

In practice, this is done by first generating nuisance parameter values from their constraint \pdf\
and by then generating \nck\ using the nuisance parameter values obtained in the first step.
Typical distributions produced by \OTH~are shown in Fig.~\ref{fig:testStatDistribExample}. 

Once $p(\qmu|\mu)$ and $p(\qmu|0)$ have been determined, \CLs\ is computed by using:
\begin{equation}
\label{eq:CLsFromDistribs}
\CLs\left(\mu\right)=\frac{\CLsb}{\CLb}=\frac{\sum\limits_{\qmu=\qmuobs}^{\infty}p(\qmu|\mu)}{\sum\limits_{\qmu=\qmuobs}^{\infty}p(\qmu|0)},
\end{equation}
where \qmuobs\ is the observed value of the test ($\qmuobs=\qmu\left(\{\nck\}=\{\nobsck\}\right)$).
The upper limit is found by searching $\mu$ such that $\CLs(\mu)$ is equal to $\alpha$ (see Eq.~\ref{eq:muUpFromCLs}).

\subsection{Computation of expected limits}
\label{sec:ExpectedLimits}

\OTH\ can also be used to compute expected limits on the signal strength, under the background only hypothesis.
Five expected limit quantiles are available: median, $-2\sigma$, $-1\sigma$, $+1\sigma$ and $+2\sigma$.
The last four ones are defined using the standard normal distribution.
Denoting these quantiles by $Z$ and the corresponding probability by $p$, one has:
\begin{equation}
Z=\Phi^{-1}\left(p\right),
\end{equation}
where $\Phi$ is the cumulative distribution function of the standard normal distribution.
The values of probabilities for the quantiles available in \OTH\ are given in Tab.~\ref{tab:ExpectedLimitProbaVsQuantile}.
\begin{table}[!htb]\centering\normalsize
\begin{tabular}{  c  c  c  c  c  c  }
\hline
$Z$  & -2 & -1 & 0 (median) & +1 & +2  \\ \hline
$p$  & 0.0228 & 0.1587 & 0.5 & 0.8413 & 0.9772 \\ \hline 
\end{tabular}
\caption{Values of probabilities associated to the quantiles available in \OTH\ for the calculation of expected limits.
\label{tab:ExpectedLimitProbaVsQuantile}}
\end{table}

Two methods are provided to compute expected limits.
In the first one, the distribution of \mup\ is determined by generating pseudo-experiments under the background-only hypothesis.
Expected limits are then given by the quantiles (as defined above) of this distribution.
In the second one, expected limits are calculated in the same way as observed ones (see Sec.~\ref{sec:InferenceMuup})
replacing the observed \CLs\ by median, $-2\sigma$, $-1\sigma$, $+1\sigma$ and $+2\sigma$ quantiles
of the \CLs\ distribution under background only hypothesis.

Both methods are equivalent from the statistical point of view but different in terms of implementation and computing time.
In most cases, the second method performs much faster than the first one and should therefore be preferred.
The first method can be used for cross-checks or if the distribution of \mup\ is needed.

However, when using the second method, quantiles are not computed from a single distribution as in the first method:
each quantile is computed in a separate job.
Therefore, due to statistical fluctuations, the expected values may not be sorted as they should
(for example, the $-2\sigma$ expected value can be higher than the $-1\sigma$ expected value).
In such cases it is recommended to increase the number of pseudo-experiments.

%% file: SoftwareDescription.tex
\section{Software description}
\label{sec:SoftDescr}
 
\subsection{Structure and prerequisites}
\label{sec:Structure}

\OTH~is written in C++ and uses the \rootcern~library~\cite{root}.
The code is separated in different classes.
Their definitions and implementations are separated into different files which are placed in the main \OTH\ directory.
Once compiled, the libraries can either be used dynamically by using a macro interpreted with the \rootcern\ interpreter
(CINT and CLING for version 5 and 6 of \rootcern, respectively), or by using a C++ program compiled into an executable with a software like gcc.
Two simple examples, working for both cases, are available in the \verb|examples| sub-directory.

As the statistical method relies on pseudo-experiments, pseudo-random number generators are used.
\rootcern\ provides in its TRandom3 class an implementation of the MT19937 Mersenne twister~\cite{Matsumoto}.
The C++ standard library in its recent C++11 standard~\cite{stroustrup} also provides various pseudo-random number engines, which generate
numbers distributed with the various probability distributions mentioned in this paper.
Therefore, if the available compiler or \rootcern\ version makes the use of the C++11 standard possible,
the user has the possibility to chose these pseudo-random engines instead of the TRandom3 class.
Thanks to relevant preprocessor directives, the portions of the code specific to the C++11 standard are ignored, either by the interpreter
or by the compiler, when its use is not possible.

If used as a compiled executable, the examples have been tested to be fully functional with gcc version $4.8.2$ ($4.3.0$) or newer,
with \rootcern\ version $5.28/00\text{b}$, when C++11 features are (are not) used.
If used as interpreted macros, the examples have been tested to be fully functional with \rootcern\ versions as old as $6.02/03$ ($5.28/00\text{b}$),
when C++11 features are (are not) used.
Older versions of gcc and \rootcern\ may also be used but have not been tested.
Other compilers than gcc could also have been chosen, but have not been considered in the installation procedure described below.

The usage instructions given below corresponds to version $2.00$ of \OTH.

\subsection{Installation and examples}
\label{sec:SoftInstall}

\OTH~is installed in three steps:
\begin{enumerate}
\item running the \verb|INSTALL| script to prepare the compilation by creating a \verb|Makefile| which structure depends on how \OTH\ is intended to be used;
\item running \verb|make| to compile the shared libraries according to the \verb|Makefile|;
\item running the \verb|setup.[c]sh| script created in the first step, to update the \verb|LD_LIBRARY_PATH| environment variable with the directory where the shared libraries are available.
\end{enumerate}
The \verb|INSTALL| shell script is in the main directory.
Using \verb|INSTALL --help| (or \verb|INSTALL -h|) displays a help detailing the different options,
which allows the user to configure the \verb|Makefile|:
\begin{itemize}
  \item \verb|--executable| (or \verb|-e|) to compile executables with gcc; if this option is not used, two \rootcern\ macros, \verb|Compile.C| and \verb|examples/load.C|,
  needed for the compilation and for loading the shared libraries, are created in addition to the \verb|Makefile|;
  \item \verb|--C++11| (or \verb|-C|) to enable the C++11 features; in this case, the script checks if the gcc or \rootcern\ versions are recent enough to do so;
  \item \verb|--permissive| to override the check of the gcc or \rootcern\ versions when using the previous option.
\end{itemize}
If no option is used, the libraries will be compiled by \rootcern\ to be used with an interpreted macro, without the C++11 features.
Once installed, \OTH~can be used from any location when opening a new shell, by running the script \verb|setup.[c]sh|.

Two examples of programs using \OTH, \verb|runLimits.C| and \verb|runSignificance.C|, are provided in the \verb|examples| sub-directory,
together with examples of input files \verb|"input1.dat"|, \verb|"input2.dat"|, \verb|"inputHistos.dat"|  and \verb|"inputHistosAlternativeSyntax.dat"|,
the syntax of which is detailed in Sec.~\ref{sec:InputFileFormat} (\verb|"input1.dat"| and \verb|"input2.dat"| illustrate the syntax for counting experiments and \verb|"inputHistos.dat"|
and \verb|"inputHistosAlternativeSyntax.dat"| for multi-bin experiments).
The example \verb|runLimits.C| illustrates the calculation of expected and observed limits, while \verb|runSignificance.C|
shows how to plot the test statistic distribution under the background only and signal plus background hypotheses and calculate the observed and expected p-values and significances.
The syntax of the main functions used in both examples is explained in Sec.~\ref{sec:Running}.

These example programs can be run 
using the following syntax:
\begin{itemize}
  \item \verb|root -l load.C 'runLimits.C("input1.dat",|\\\verb|  "input2.dat")'|\\when used as a \rootcern\ macro;
  \item \verb|./runLimits.exe --files input1.dat|\\\verb|  input2.dat|\\when used as an executable.
\end{itemize}
Here, files for counting experiments are provided. Files for multi-bin experiments are provided using exactly the same syntax (\OTH{} automatically recognises what type of input files are provided by the user and applies the right treatment in each case). 
The input files don't have to be of the same type. For example, it is possible to provide one file using the counting experiment syntax and another file using the multi-bin experiment syntax :
 \begin{itemize}
  \item \verb|root -l load.C 'runLimits.C("input1.dat",|\\\verb|  "inputHistosAlternativeSyntax.dat")'| ;
  \item \verb|./runLimits.exe --files input1.dat|\\\verb|  inputHistosAlternativeSyntax.dat|
\end{itemize}

In these examples, preprocessor directives are used to allow their use in both modes, or to enable the C++11 features.
In the first mode, these examples can also be run from any location, if the \verb|load.C| macro written to load the shared libraries is run before.
In the second mode, the executable can be moved to any location; furthermore, any program located in the \verb|examples| sub-directory would be compiled
when running the \verb|make| command, as long as its name is of the form \verb|run*.C|.

\subsection{Input file format}
\label{sec:InputFileFormat}

The input file format depends on the type of experiment. 
The format for counting experiments is described in Sec.~\ref{sec:InputFileFormatCounting} and the one for multi-bin experiments is described in Sec.~\ref{sec:InputFileFormatShape}.
In both cases, users must create one input file per channel.

\subsubsection{Counting experiment}
\label{sec:InputFileFormatCounting}

The general structure of the files is shown if Fig.~\ref{fig:generalLayoutInputFile}.

\begin{figure}[!htb]
\centering
\begin{BVerbatim}
+sig <sig_name> <yield> <stat>
.syst <sys1_name> <up> <down>
.syst <sys2_name> <up> <down>

+bg <bkg1_name> <yield> <stat>
.syst <sys1_name> <up> <down>
.syst <sys2_name> <up> <down>

+bg <bkg1_name> <yield> <stat>
.syst <sys1_name> <up> <down>
.syst <sys2_name> <up> <down>

# ...
# add as many backgrounds as needed
# ...

+data <yield>
\end{BVerbatim}
\caption{General structure of \OTH's input files for counting experiments. \label{fig:generalLayoutInputFile}}
\end{figure}

The block starting with the \verb|+sig| tag defines the signal sample.   
Blocks starting with the  \verb|+bg| tag define the background samples.   
Finally, the observation is defined with the \verb|+data| tag.
Systematic uncertainties are declared using the \verb|.syst| tag. 
Fields \verb|<name>| define the names for signal, backgrounds and systematic uncertainties.
Fields \verb|<yield>| define the nominal yields for signal and backgrounds and the observed yield for data.
Fields \verb|<stat>| define the absolute statistical uncertainty for signal and backgrounds.
Fields \verb|<up>| and \verb|<down>| define \fup~and \fdown~for each systematic uncertainty.
When systematics for different samples (signal or backgrounds) or different channels have the same name
they are supposed 100\% correlated (i.e. they are described by a unique nuisance parameter $\eta_j$).
Otherwise they are treated as uncorrelated.
Lines starting with \verb|#| are not interpreted.
The order in which the signal, background and data block appear does not matter. 

\latex{} tables summarizing yields and uncertainties can be produced (see Sec.~\ref{sec:TeXTablesProduction} for instructions on how to do this). 
By default, the sample and systematic names used in these tables are the ones specified in the input files in the \verb|<name>| fields. 
The user can change background sample names by adding \verb|.nameLaTeX| tags in background sample blocks.   
Systematic names can be changed by creating a dictionnary file mapping names as specified in the \verb|<name>| fields to names used for \latex{} tables.
Channel names used for \latex{} tables can also be specified by adding \verb|+nameLaTeX| tags in the input files. 

Fig.~\ref{fig:ExampleFilesTwoChannels} and~\ref{fig:ExampleFilesTwoChannelsDict} show an example of what could be the two files
in an analysis combining two channels and the dictionnary file defining systematic names for \latex{} tables.

\begin{figure}[!htb]
\centering
\begin{BVerbatim}
+nameLaTeX $e\mu$

+bg Bkg1 0.8 0.1
.nameLaTeX $t\bar{t}$ 
.syst Syst1 -0.05 0.12
.syst Syst2 0.04 -0.04

+sig Sig 2.5 0.6
.syst Syst1 0.21 -0.13

+data 1
\end{BVerbatim}
\hspace{10pt}
\begin{BVerbatim}
+nameLaTeX $\mu\mu$

+bg Bkg2 2.3 0.4
.nameLaTeX $t\bar{t}+W/Z$ 
.syst Syst3 0.01 0.01

+sig Sig 2.8 1.1
.syst Syst1 0.05 -0.13
.syst Syst4 -0.02 -0.09

+data 3
\end{BVerbatim}
\caption{Example of input text files for a two channels combination. \label{fig:ExampleFilesTwoChannels}}
\end{figure}

\begin{figure}[!htb]
\centering
\begin{BVerbatim}
Syst1 JES
Syst2 $b$-tagging
Syst3 $E_{T}^{\rm miss}$
Syst4 norm
\end{BVerbatim}
\caption{Example of dictionnary file defining systematic names for \latex{} tables. \label{fig:ExampleFilesTwoChannelsDict}}
\end{figure}

In this example, all backgrounds and signals have non-zero statistical uncertainty.
The total number of systematic uncertainties (and thus of nuisance parameters $\eta_j$) is 4.
The one called \verb|Syst1| in the input files (\verb|JES| in the \latex{} tables) affects the background yield
of the first channel and signal yields of the two channels.
\latex{} tables produced by \OTH{} for this example are shown in Tab.~\ref{tab:ExampleChannel12yields}, \ref{tab:ExampleChannel12yieldsWithToys}, \ref{tab:ExampleChannel1Systs} and~\ref{tab:ExampleChannel2Systs}, including captions that are generated automatically.
Tab.~\ref{tab:ExampleChannel12yields} gives the observed yields in all channels. 
Expected nominal yields together with their statistical and total systematic uncertainties for all samples are also given.
Expected nominal yields and statistical uncertainties are those given by the user in the input files. 
The total systematic uncertainties are given by the squared sum of the systematic uncertainties provided by the user in the input files.
Table.~\ref{tab:ExampleChannel12yieldsWithToys} also gives observed and expected yields with uncertainties for all samples in all channels. 
However, expected yields and their uncertainties are now computed from quantiles of the marginal distribution of the yield (where the marginalization is performed over all nuisance parameters, statistical and systematic).
The expected yields are given by the medians of the marginal distributions and the intervals between the lower and upper uncertainty are the $1\sigma$ confidence level interval.

\begin{table}[!htb]\begin{center}
\begin{tabular}{l *{2}{c}}
\hline\hline
Sample & $e\mu$ & $\mu\mu$ \\
\hline\hline
$t\bar{t}$  & $0.80 \pm 0.10 ^{+0.10}_{-0.05}$ &  ---  \\
\hline
$t\bar{t}+W/Z$  &  ---  & $2.30 \pm 0.40 ^{+0.02}_{-0.00}$ \\
\hline
\hline
Total bkg. & $0.80 \pm 0.10 ^{+0.10}_{-0.05}$ & $2.30 \pm 0.40 ^{+0.02}_{-0.00}$ \\
\hline\hline
Data & 1 & 3 \\
\hline\hline
Signal & $2.50 \pm 0.60 ^{+0.53}_{-0.33}$ & $2.8 \pm 1.1 ^{+0.1}_{-0.4}$ \\
\hline\hline
\end{tabular}
\caption{Observed yields and nominal expected yields. For each nominal expected yield, the first quoted uncertainty represent the statistical uncertainty, while the second is an approximation of the total systematic uncertainty, without taking into account the correlations between them.\label{tab:ExampleChannel12yields}}
\end{center}\end{table}

\begin{table}[!htb]\begin{center}
\begin{tabular}{l *{2}{c}}
\hline\hline
Sample & $e\mu$ & $\mu\mu$ \\
\hline\hline
$t\bar{t}$  & $0.81 ^{+0.14}_{-0.12}$ &  ---  \\
\hline
$t\bar{t}+W/Z$  &  ---  & $2.28 ^{+0.42}_{-0.38}$ \\
\hline
\hline
Total bkg. & $0.81 ^{+0.14}_{-0.12}$ & $2.28 ^{+0.42}_{-0.38}$ \\
\hline\hline
Data & 1 & 3 \\
\hline\hline
Signal & $2.47 ^{+0.84}_{-0.63}$ & $2.4 ^{+1.2}_{-0.9}$ \\
\hline\hline
\end{tabular}
\caption{Observed yields and median expected yields. For each median expected yield, derived with pseudo-experiments, the quoted uncertainty is a combination of the statistical and systematic uncertainties, taking into account the correlations between systematics.\label{tab:ExampleChannel12yieldsWithToys}}
\end{center}\end{table} 

\begin{table}[!htb]\begin{center}
\begin{tabular}{l *{2}{c}}
\hline\hline
Uncertainty & $t\bar{t}$  & Signal \\
\hline\hline
JES & $_{+12.00}^{-5.00}$ & $_{-13.00}^{+21.00}$ \\
\hline
$b$-tagging & $\pm 4.00$ & --- \\
\hline
$E_{T}^{\rm miss}$ & --- & --- \\
\hline
norm & --- & --- \\
\hline
\hline
Total & $_{-6.40}^{+12.65}$ & $_{-13.00}^{+21.00}$ \\
\hline\hline
\end{tabular}
\caption{List of relative systematic uncertainties (in \%) for channel $e\mu$.\label{tab:ExampleChannel1Systs}}
\end{center}\end{table} 

\begin{table}[!htb]\begin{center}
\begin{tabular}{l *{2}{c}}
\hline\hline
Uncertainty & $t\bar{t}+W/Z$  & Signal \\
\hline\hline
JES & --- & $_{-13.00}^{+5.00}$ \\
\hline
$b$-tagging & --- & --- \\
\hline
$E_{T}^{\rm miss}$ & $_{+1.00}^{+1.00}$ & --- \\
\hline
norm & --- & $_{-9.00}^{-2.00}$ \\
\hline
\hline
Total & $_{+0.00}^{+1.00}$ & $_{-15.81}^{+5.00}$ \\
\hline\hline
\end{tabular}
\caption{List of relative systematic uncertainties (in \%) for channel $\mu\mu$. \label{tab:ExampleChannel2Systs}}
\end{center}\end{table}

\subsubsection{Multi-bin experiment}
\label{sec:InputFileFormatShape}

Two syntaxes can be used. Both are similar to that for counting experiments, except that \rootcern~files containing histograms are provided instead of numbers. Only differences with respect to the syntax described in Sec.~\ref{sec:InputFileFormatCounting} are therefore described.

\paragraph{Syntax 1:}

This syntax can be used only if all \rootcern~files are in the same directory and all histograms inside them have the same name. The general structure of the files is shown in Fig.~\ref{fig:generalLayoutInputFileShapeSyntax1}.
An example of such a file is \verb|"inputHistos.dat"| in the \verb|examples| sub-directory. 

\begin{figure}[!htb]
\centering
\begin{BVerbatim}
+setup
.directory <dir>
.histoName <hname>

+sig <sig_name> <root-file> 
.syst <sys1_name> <root-file up> <root-file down>
.syst <sys2_name> <root-file up> <root-file down>

+bg <bkg1_name> <root-file> 
.syst <sys1_name> <root-file up> <root-file down>
.syst <sys2_name> <root-file up> <root-file down>

+bg <bkg2_name> <root-file> 
.syst <sys1_name> <root-file up> <root-file down>
.syst <sys2_name> <root-file up> <root-file down>

# ...
# add as many backgrounds as needed
# ...

+data <root file> 
\end{BVerbatim}
\caption{General structure of \OTH's input files for multi-bin experiments using syntax 1. \label{fig:generalLayoutInputFileShapeSyntax1}}
\end{figure}

The block starting with the \verb|+setup| tag defines the fields that are common to all samples declared in the file (signal, backgrounds and data). 
Field \verb|<dir>| defines the directory containing all the \rootcern{} files.
Field \verb|<hname>| defines the name of the histograms inside the \rootcern{} files.
Fields \verb|<root-file>| define the names of the \rootcern~files located in the \verb|<dir>| directory 
containing the nominal histograms.
Bin uncertainties are used for absolute statistical uncertainties. 
Therefore, care must be taken in the case of histograms filled with weighted entries.
Fields \verb|<root-file up>| and \verb|<root-file down>| define the names of the \rootcern~files located 
in the \verb|<dir>| directory containing the varied histograms when systematic uncertainties are shifted by 
$+1\sigma$ and $-1\sigma$ respectively.
These histograms must contain yields in each bin (that is, as opposed to the counting experiment case, yields after systematic 
variation are provided rather than relative variations).
In addition, bin uncertainties for systematic histograms are not used and can be set to $0$ or any other value. 

\paragraph{Syntax 2:}

This syntax can be used in all cases, even if all \rootcern~files aren't in the same directory and all histograms inside them don't have the same name. The general structure of the files is shown in Fig.~\ref{fig:generalLayoutInputFileShapeSyntax2}. An example of such file is \verb|"inputHistosAlternativeSyntax.dat"| in the \verb|examples| sub-directory. 

\begin{figure*}[!htb]
\centering
\begin{BVerbatim}
+sig <sig_name> <root-file>(<hname>) 
.syst <sys1_name> <root-file up>(<hname>) <root-file down>(<hname>) 
.syst <sys2_name> <root-file up>(<hname>) <root-file down>(<hname>) 

+bg <bkg1_name> <root-file>(<hname>)  
.syst <sys1_name> <root-file up>(<hname>) <root-file down>(<hname>) 
.syst <sys2_name> <root-file up>(<hname>) <root-file down>(<hname>) 

+bg <bkg2_name> <root-file>(<hname>)  
.syst <sys1_name> <root-file up>(<hname>) <root-file down>(<hname>) 
.syst <sys2_name> <root-file up>(<hname>) <root-file down>(<hname>) 

# ...
# add as many backgrounds as needed
# ...

+data <root file>(<hname>) 
\end{BVerbatim}
\caption{General structure of \OTH's input files for multi-bin experiments using syntax 2. \label{fig:generalLayoutInputFileShapeSyntax2}}
\end{figure*}

Fields \verb|<root-file>|,  \verb|<root-file up>| and \verb|<root-file down>| define the 
full path names of the \rootcern{} files containing the nominal and varied histograms.
Fields \verb|<hname>| define the names of the histograms inside the \rootcern{} files.
They must be between parentheses with no blank space between the left parenthesis and the \rootcern{} file name.

\subsection{Running the software}
\label{sec:Running}

\subsubsection{Configuration}
\label{sec:OTHInstance}

In order to use the software, users need to load the \OTH\ library, instantiate an object of type \verb|OpTHyLiC| and add channels.
This can be done interactively in a \rootcern\ session as follows:
\begin{verbatim}
  gSystem->Load("OpTHyLiC_C");
  OpTHyLiC oth(OTH::SystMclimit,
    OTH::StatNormal,OTH::TR3,0,
    OTH::CombAutomatic);
  oth.addChannel("channel 1 name",file1);
  oth.addChannel("channel 2 name",file2);
  ...
\end{verbatim}

The constructor requires at least two parameters.
The first one is the interpolation/extrapolation method: the possible values are \verb|OTH::SystLinear|, \verb|OTH::SystExpo|,
\verb|OTH::SystPolyexpo|, and \verb|OTH::SystMclimit|, corresponding to the four possibilities described in Sec.~\ref{sec:SystUncertTreatment}.
The second one is the type of constraint for statistical uncertainties: the possible values are \verb|OTH::StatNormal|, \verb|OTH::StatLogN|,
\verb|OTH::StatGammaHyper|, \verb|OTH::StatGammaUni|, and \verb|OTH::StatGammaJeffreys|,
corresponding to the five possibilities described in Sec.~\ref{sec:StatUncertTreatment}.

In addition, three optional parameters can be provided.
The first one is the type of pseudo-random number engine: the possible values are \verb|OTH::TR3| (used by default), and, if C++11 features are available,
\verb|OTH::STD_| followed by the name of one of the nine engines implemented in the C++11 standard library
(\verb|mt19937|, \verb|mt19937_64|, \verb|minstd_rand|, \verb|minstd_rand0|, \verb|ranlux24_base|, \verb|ranlux48_base|, \verb|ranlux24|, \verb|ranlux48|,  \verb|knuth_b|).
The second one is the seed of this engine; the default value \verb|0| has the effect of setting a randomly generated seed.
The third one is the chosen solution for the combination of systematic uncertainties, as described in Sec.~\ref{sec:SystUncertTreatment}:
the possible values are \verb|OTH::CombAdditive|, \verb|OTH::CombMultiplicative|, or \verb|OTH::CombAutomatic| (used by default),
the latter requesting additive (multiplicative) combination when the \verb|OTH::SystLinear| or \verb|OTH::SystMclimit|
(\verb|OTH::SystExpo| or \verb|OTH::SystPolyexpo|) options is used for the interpolation/extrapolation method.

The \verb|addChannel| member function takes three parameters. 
The first two ones are of \verb|std::string| type and correspond to the name of the channel and the name of the input file (see Sec.~\ref{sec:InputFileFormat}) respectively.
The third one, used only in the case of multi-bin experiments, is a boolean set by default to \verb|true|. 
In the case of multi-bin experiments, \OTH{} converts the input file and ROOT files provided by the user into multiple counting experiment files (written using the syntax described in Sec.~\ref{sec:InputFileFormatCounting}). 
More precisely, \OTH{} creates one counting experiment file per bin of discriminating variable (only bins containing either a non-zero yield or a non-zero statistical uncertainty for the signal and total background are considered).
If the third parameter is set to \verb|false|, counting experiment files created by \OTH{} are dumped on disk, in the same directory as the input file. 
Their names are of the form \verb|fileName_bin|$\theta$\verb|.ext|, where \verb|fileName| is the name of the input file, \verb|ext| its extension and $\theta$ is the bin index.
These files can be useful for debugging or for running \OTH{} over a subset of bins.

As mentioned in Sec.~\ref{sec:StatModel}, for multi-bin experiments the histogram bins are treated in the same way as channels in counting experiments.
Therefore, in the rest of this paper the term ``channel'' refers in such case to a bin of the discriminating variable distribution in one channel,
reflecting the software structure.

\subsubsection{\latex{} tables production}
\label{sec:TeXTablesProduction}

\latex{} tables are produced as follows:
\begin{verbatim}
  ofstream ofs(filename);
  int precision = -1;
  int NpseudoExps = 1000000;
  oth.createInputYieldTable(ofs,precision);
  oth.createGeneratedYieldTable(ofs,
    precision,NpseudoExps);
  oth.createSysteTables(ofs,"systDict.txt",
    precision);
\end{verbatim}
These three methods take as first parameter an output stream, which can be the standard output (\verb|std::cout|). 

The functions \verb|createInputYieldTable| and \verb|createGeneratedYieldTable| produce tables of expected and observed yields.
In the first case, the total uncertainty for each process is evaluated by summing in quadrature all uncertainties
without taking into account the correlations, while in the second case the total uncertainty is assessed by using pseudo-experiments (the number of pseudo-experiments is given as the facultative
third parameter).
An example of such tables are shown in Tab.~\ref{tab:ExampleChannel12yields} and \ref{tab:ExampleChannel12yieldsWithToys}. The precision of the
numbers in these tables can be adjusted to a fixed value (given as the second parameter) or to 
be automatically adjusted depending on the size of the uncertainties (when the precision 
parameter is set to -1, which is the default value).

The function \verb|createSysteTables| creates one table for each channel which summarises all the systematic uncertainties. 
Examples of such tables are shown in Tab.~\ref{tab:ExampleChannel1Systs} and \ref{tab:ExampleChannel2Systs}.
The second parameter of \verb|createSysteTables| is a string specifying the dictionary file.
An example of such a file is given in Fig.~\ref{fig:ExampleFilesTwoChannelsDict}.

\subsubsection{Observed limit computation}
\label{sec:observedLimitComputation}

Observed limits are computed as follows:
\begin{verbatim}
  double cls;
  double limit=oth.sigStrengthExclusion(
    OTH::LimObserved,nbExp,cls);
\end{verbatim}
The first parameter of function \verb|sigStrengthExclusion| defines the limit type (here observed). 
Other types can be used for expected limits (see Sec.~\ref{sec:expectedLimitComputation}).
The second parameter (\verb|nbExp|) is the number of pseudo-experiments
and the third parameter the final \CLs~value (corresponding to the exclusion).
An optional fourth parameter may be provided and corresponds to a hint of the observed limit. This value
does not affect the result of the computation but only its speed: giving a hint that is rather close to the
final value would most probably speed up the process. Finally, an optional fifth parameter may be used
to determine which method to use for the computation. The default value (\verb|OTH::MethDichotomy|) is the
most accurate one. The other possibility is to use an extrapolation method (\verb|OTH::MethExtrapol|) that
is faster but sometimes less accurate.

By default, the computed limits are for a 95\% confidence level. It is possible to modify
this confidence level as follows:
\begin{verbatim}
  oth.setConfLevel(0.9);
\end{verbatim}
for example for a 90\% confidence level.

\subsubsection{Expected limits computation}
\label{sec:expectedLimitComputation}

Two methods are provided to compute expected limits as described in Sec.~\ref{sec:ExpectedLimits}.

With the main method,  a single expected (median, $-2\sigma$, $-1\sigma$, $+1\sigma$ or $+2\sigma$) limit is computed as the observed one (see Sec.~\ref{sec:observedLimitComputation}) changing the first parameter to \verb|OTH::LimExpectedM2sig|,  \verb|OTH::LimExpectedM1sig|,  \verb|OTH::LimExpectedMed|, \verb|OTH::LimExpectedP1sig| or \verb|OTH::LimExpectedP2sig| depending on the limit type.
For example, the expected median limit is computed using:
\begin{verbatim}
  double limit=oth.sigStrengthExclusion(
    OTH::LimExpectedMed,nbExp,cls);
\end{verbatim}
Like in Sec.~\ref{sec:observedLimitComputation}, it is possible to provide a hint of the expected limit and
to use the extrapolation method when speed is preferred to accuracy.

With the alternative method, all expected (median, $-2\sigma$, $-1\sigma$, $+1\sigma$ and $+2\sigma$) limits are computed at the same time using:
\begin{verbatim}
  oth.expectedSigStrengthExclusion(nbMu,nbExp);
\end{verbatim}
where \verb|nbMu| and \verb|nbExp| are the number of entries in the \mup~distribution (see Sec.~\ref{sec:ExpectedLimits})
and the number of pseudo-experiments, respectively.

\subsubsection{Scan of \CLs~as a function of the signal strength}
It is possible to perform a scan of the \CLs{} value for several values of the signal strength, using:
\begin{verbatim}
  oth.scanCLsVsMu(min,max,steps,nbExp,
    OTH::LimExpectedMed);
  oth.getCLsVsMu()->Draw("alp");
\end{verbatim}
The first three parameters define the signal strength range to be scanned, \verb|steps| being the number of
values to be scanned. The parameter \verb|nbExp| is the number of pseudo-experiments to compute each \CLs.
The last parameter defines the type of observed value to be used, within the list \verb|OTH::LimExpectedM2sig|,  \verb|OTH::LimExpectedM1sig|,  \verb|OTH::LimExpectedMed|, \verb|OTH::LimExpectedP1sig| or \verb|OTH::LimExpectedP2sig| and \verb|OTH::LimObserved|.

By fitting the returned graph, the user may compute the corresponding limit, alternatively to the
methods presented in Sec.~\ref{sec:observedLimitComputation} and~\ref{sec:expectedLimitComputation}.

\subsubsection{Significance computation}
\label{sec:significanceComputation}

In addition to observed and expected upper limits, \OTH{} can be used to compute the significance of an observation using:
\begin{verbatim}
  std::pair<double, double> r
    = oth.significance(type,nbExp,mu);
\end{verbatim}
where \verb|type| is the type of significance, \verb|nbExp| the number of pseudo-experiments and \verb|mu| the signal strength (set by default to $1$ if not specified). 
The possible values for \verb|type| are 
\verb|OTH::SignifObserved|, 
\verb|OTH::SignifExpectedP2sig|, 
\verb|OTH::SignifExpectedP1sig|, 
\verb|OTH::SignifExpectedMed|, 
\verb|OTH::SignifExpectedM1sig|, 
\verb|OTH::SignifExpectedM2sig|.
If \verb|OTH::SignifObserved| is used, the observed significance is computed. Otherwise, expected significances under the signal plus background hypothesis are computed. 

The \verb|significance| method returns a \verb|std::pair<double, double>|. The first element of the pair corresponds to the \pval{} given by:
\begin{equation}
\pval=\sum\limits_{\qmu=0}^{\qmuobs}p(\qmu|0)
\end{equation}
and the second one to the significance $z$ in units of normal distribution standard deviation. It is given by:
\begin{equation}
\pval=\displaystyle\int_{z}^{\infty}\frac{1}{\sqrt{2\pi}}e^{-\frac{x^2}{2}}\dd x
\end{equation}

\subsubsection{Access to histograms}
\label{sec:AccessToHistos}
During the limit computation, several histograms are produced. After the computation, some
of these histograms are available.

The final \qmu{} distributions can be accessed using the following commands:
\begin{verbatim}
  oth.getHistoLLRb()->Draw();
  oth.getHistoLLRsb()->Draw("same");
\end{verbatim}
In the case of the alternative expected limit computation, the \qmu{} distributions are not
accessible but two other histograms are available:
\begin{verbatim}
  oth.getDistrExpMu()->Draw();
  oth.getDistrCLs()->Draw();
\end{verbatim}
where the first one is the expected distribution of \mup{} from which the expected limits
are derived and the last one is the full distribution of computed \CLs{} (that should peak at 
1-confidence level).

Other interesting distributions are channel dependent, therefore the user must first access
a specific channel. There are two ways to do so:
\begin{itemize}
\item \verb|oth.getChannel("name")| where \verb|name| is the name of the channel,
\item \verb|oth.getChannel(index)| where \verb|index| is the index of the channel returned by the \verb|addChannel| function.
\end{itemize}
In the case of multi-bin experiment, \verb|"name"| should be of the form \verb|channelName_bin|$\theta$, where \verb|channelName|
is the name of the channel specified by the user (first parameter of the \verb|addChannel| method) and $\theta$ is the bin index.
The final \qmu{} distributions for a single channel can be accessed using the following commands:
\begin{verbatim}
  oth.getChannel(index)
    ->getHistoLLRb()->Draw();
  oth.getChannel(index)
    ->getHistoLLRsb()->Draw("same");
\end{verbatim}
as well as the distributions of number of events:
\begin{verbatim}
  oth.getChannel(index)
    ->getHisto(OTH::Channel::hDistrBg)
    ->Draw();
  oth.getChannel(index)
    ->getHisto(OTH::Channel::hDistrSB)
    ->Draw("same");
\end{verbatim}
Other interesting histograms are the distributions of the systematic uncertainties, for example:
\begin{verbatim}
  oth.getChannel(index)
    ->getSigSystDistr("Syst1")
    ->Draw();
  oth.getChannel(index)
    ->getBkgSystDistr("Bkg","Syst2")
    ->Draw();
\end{verbatim}
where the last parameter is the name of the systematic uncertainty and the first parameter of
the \verb|getBkgSystDistr| function is the name of the background.

\subsubsection{Single channel computations}
To compute the limits for a single channel, not combining with the others, 
the following function is available:
\begin{verbatim}
  double limit=oth.getChannel(index)
    ->sigStrengthExclusion(OTH::LimObserved,
    nbExp,cls);
\end{verbatim}
After this call, channel dependent histograms (as described in the previous section) are
available.

For the alternative expected limit computation, the following function may be invoked:
\begin{verbatim}
  double limit=oth.getChannel(index)
    ->expectedSigStrengthExclusion(nbMu,nbExp);
\end{verbatim}
In this case, all previous histograms are available as well as a few more:
\begin{verbatim}
  oth.getChannel(index)
    ->getDistrExpMu()->Draw();
  oth.getChannel(index)
    ->getExpMuVsObs()->Draw("alp");
  oth.getChannel(index)
    ->getDistrCLs()->Draw();
\end{verbatim}
where the second one is the expected \mup{} as a function of the number of observed
events.

It is also possible to study the distribution of the yields after the statistical and systematic
variations. In order to generate these distributions, the following function must be called:
\begin{verbatim}
  oth.getChannel(index)
    ->generateDistrYield(nbExp);
\end{verbatim}
then, the generated distributions are available:
\begin{verbatim}
  oth.getChannel(index)
    ->getSigYieldDistr()
    ->Draw();
  oth.getChannel(index)
    ->getBkgYieldDistr("Bkg")
    ->Draw();
  oth.getChannel(index)
    ->getHisto(OTH::Channel::hYieldBg)
    ->Draw();
\end{verbatim}
where the first function gives the distribution for the signal, the second one for a single background 
sample, and the third one is the distribution for the total background. These distributions are also
available after a call to \verb|createGeneratedYieldTable|.

\subsubsection{Test statistic distribution and derived quantities}

The test statistic distributions under background and signal plus background hypotheses for a desired value of the signal strength $\mu$ can be computed as follows:
\begin{verbatim}
  oth.setSigStrength(mu);
  oth.generateDistrLLR(nbExp);
\end{verbatim}
where \verb|mu| is the desired $\mu$ value and \verb|nbExp| is the number of pseudo-experiments. Once this is done, distributions can be accessed as described in Sec.~\ref{sec:AccessToHistos}. The \pval{} of the observation under the background hypothesis and the \CLs\ can be computed as follows:
\begin{verbatim}
  double pvalue=oth.pValueData();
  double cls=oth.computeCLsData();
\end{verbatim}

The \pval{} under the background hypothesis and the \CLs\ for a single channel can be computed using:
\begin{verbatim}
  double pvalue=oth.getChannel(index)
    ->pValue(nObs);
  double cls=oth.getChannel(index)
    ->computeCLs(nObs);
\end{verbatim}
where \verb|nObs| is the observed yield. For a single channel, the user can also compute the excluded yield for the chosen $\mu$ value using:
\begin{verbatim}
  int yield=oth.getChannel(index)
    ->findObsExclusion();
\end{verbatim}
 
\subsection{Improvement of the software performance}
\label{sec:Performance}

Thanks to the absence of profiling, the computation of the pseudo-experiments could be optimised
to minimise the number of random numbers that need to be drawn.
Moreover, when searching for the signal strength that corresponds to a given \CLs, it is possible
to use a simple extrapolation from two pairs of (\CLs,$\mu$) values, that is much faster than a real scan,
but has a larger uncertainty on the result.
For a more precise but slower result, dichotomy is used, optimised by taking into account the logarithmic
dependence between the \CLs\ and the signal strength.

As will be shown in Sec.~\ref{sec:ValidationWithUncertainties}, even with the dichotomy choice, observed and expected limits are in general computed very fast with \OTH~compared to other software tools. 
Speedups well over a factor of 10 have for example been achieved with respect to \mclimit.

%% file: SoftwareValidation.tex
\section{Software validation}
\label{sec:SoftValid}

The computation of upper limits and observation significances with \OTH~involves many calculations of different types:
generation of random numbers, calculation of \pval s (\CLsb~and \CLb), scanning over $\mu$ values to solve Eq. \ref{eq:muUpFromCLs},
combination of channels, interpolation and extrapolation of systematic uncertainties, marginalization of statistical and systematic uncertainties,
calculation of quantiles for expected limits, etc.
Several studies have been performed in order to validate these calculations.
\OTH~is compared either to a theoretical solution in situations where such a solution exists,
to Bayesian calculations in situations where they are equivalent to the hybrid \CLs~under interest or to results
obtained with the \mclimit~software which is equivalent to \OTH~when specific choices for the interpolation/extrapolation of systematics
and constraints for statistical uncertainties are made.
These studies are summarized in the following sections.

\subsection{Validation of calculations without uncertainties}

\OTH~has first been validated in the simplest situation where signal and background yields have no uncertainties
(neither systematic nor statistical), both in the single and multiple channels cases.

In the single channel case, an analytical solution for \mup~exists.
Indeed, \CLsb~and \CLb~are given by:
\begin{equation}
\label{eq:defCLsb}
\begin{array}{rl}
\CLsb=&\sum\limits_{\n=0}^{\nobs}\frac{\left(\mu\snom+\bnom\right)^{\n}}{\n!} e^{-\left(\mu\snom+\bnom\right)}\\
&\\
     =&1-F_{\chi^2}\left(2\left(\mu\snom+\bnom\right);2\left(\nobs+1\right)\right)\\
\end{array}
\end{equation}
and:
\begin{equation}
\label{eq:defCLb}
\CLb=\sum\limits_{\n=0}^{\nobs}\frac{\left(\bnom\right)^{\n}}{\n!} e^{-\bnom}=1-F_{\chi^2}\left(2\bnom;2\left(\nobs+1\right)\right),
\end{equation}
where $F_{\chi^2}\left(x;d\right)$ is the cumulative distribution function
of the chi-squared distribution with $d$ degrees of freedom at $x$.
Eq.~\ref{eq:muUpFromCLs} then yields:
\begin{equation}
\label{eq:muUpAnalyticalResultWoUncert}
\mup=\frac{0.5\times F_{\chi^{2}}^{-1}\left(1-\alpha\left[1-F_{\chi^{2}}\left(2\bnom;2\left(\nobs+1\right)\right)\right];2\left(\nobs+1\right)\right)-\bnom}{\snom}
\end{equation}

\begin{figure}[!htb]
\begin{center}
\includegraphics[scale=0.45]{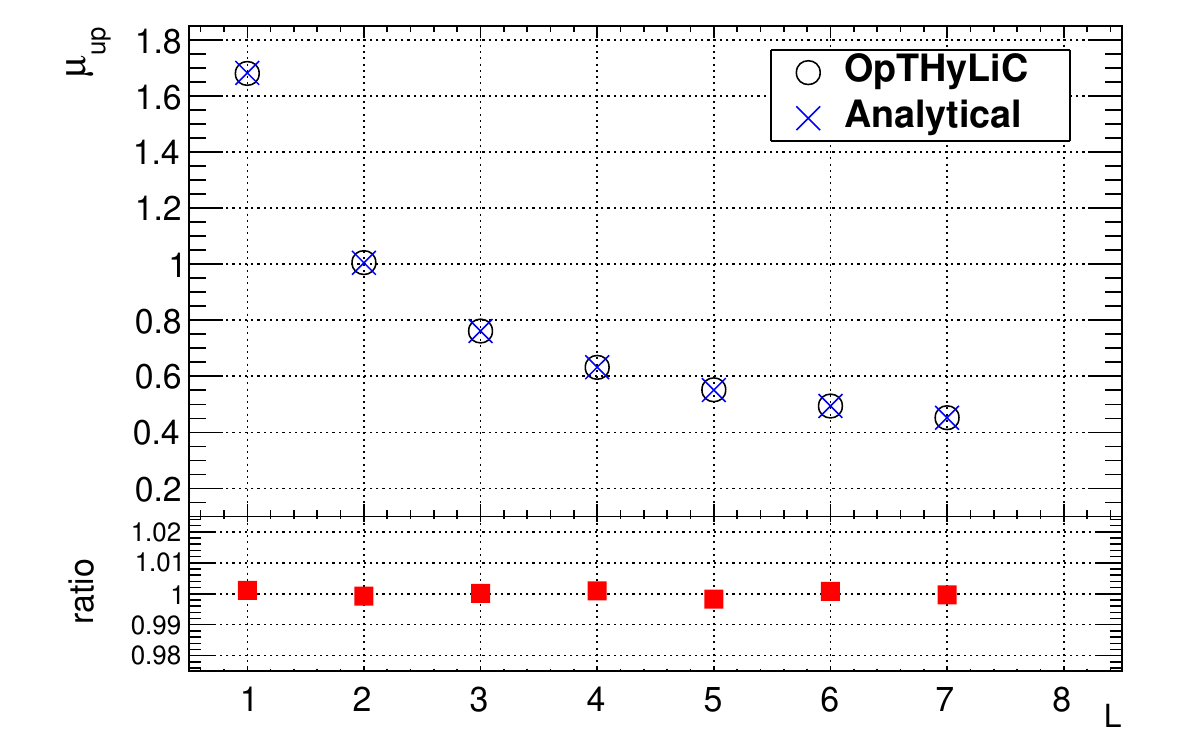}
\vspace{-2pt}
\caption{Upper limit \mup~as a function of $L$ computed with \OTH~and from the analytical result (Eq. \ref{eq:muUpAnalyticalResultWoUncert}) for $\bnom=0.82\times L$, $\snom=2.49\times L$ and $\nobs=1\times L$.\label{fig:ExampleValidNoUncertVsAnalytical}}
\end{center}
\end{figure}

Fig. \ref{fig:ExampleValidNoUncertVsAnalytical} shows a comparison between this analytical
result and \OTH~for $\bnom=0.82\times L$, $\snom=2.49\times L$ and $\nobs=1\times L$, with $L=1,\hdots,7$.
Excellent agreement is found.
Other tests have been performed with other values of $\bnom$, $\snom$ and $\nobs$, always leading to the same conclusion.

In the multiple channels case, two validations were made.
In the first one, upper limits calculated with several channels were compared to upper limits
calculated with a single channel in situations where the two are expected to give identical results.
Such situations occur when yields in the various channels are related to each other by a simple multiplicative factor.
For example, the same limits should be obtained in these two cases: 
\begin{itemize}
  \item \nochan~channels each with:
  \begin{itemize}
    \item background yield=$\bnom/\nochan$
    \item signal yield=$\snom/\nochan$
    \item observed yield=$\nobs/\nochan$
  \end{itemize}
  \item a single channel with:
  \begin{itemize}
    \item background yield=$\bnom$
    \item signal yield=$\snom$
    \item observed yield=$\nobs$
  \end{itemize}
\end{itemize}

\begin{figure}[!htb]
\begin{center}
\includegraphics[scale=0.45]{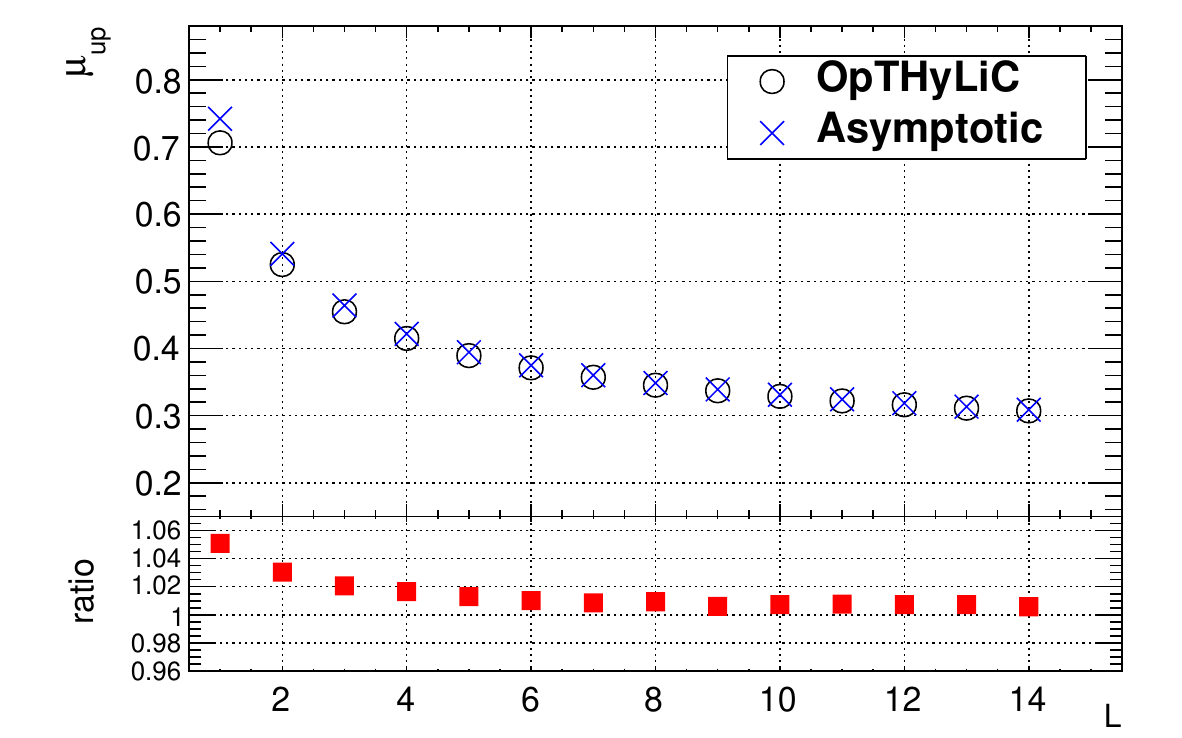}
\vspace{-2pt}
\caption{Upper limit \mup~as a function of $L$ computed with \OTH~and from the asymptotic result (Eq. \ref{eq:CLsMultipleChannelsWoUncertAsympt}) for the three channels example defined in the text. \label{fig:ThreeChannelExampleAsympt}}
\end{center}
\end{figure}

It has been checked for several values of \snom, \bnom, \nobs~and \nochan~that \OTH~indeed finds the same limits in both cases.
In the second one, the general case where yields in the various channels can't be related to each other by a simple multiplicative factor has been considered.
It can be seen from Eq. \ref{eq:TestStatAllChannels} and Eq. \ref{eq:TestStatChannelc} that the upper limit can be computed using the following test:
\begin{equation}
\n_{\text{eff}}=\sum\limits_c \nc\beta_c\quad\text{with}\quad\beta_c=\ln\frac{\mu\scnom+\bcnom}{\bcnom}.
\end{equation}
In the asymptotic limit, $\nc$ is normally distributed. Thus:
\begin{equation}
\n_{\text{eff}}\sim {\cal N}\left(\sum\limits_c\beta_c\left(\mu\scnom+\bcnom\right),\sum\limits_c\beta_c^2\left(\mu\scnom+\bcnom\right)\right),
\end{equation}
under the signal plus background hypothesis and:
\begin{equation}
\n_{\text{eff}}\sim {\cal N}\left(\sum\limits_c\beta_c\bcnom,\sum\limits_c\beta_c^2\bcnom\right),
\end{equation}
under the background only hypothesis (${\cal N}\left(a,b\right)$ is the normal distribution with mean $a$ and variance $b$).
\CLs~is therefore given by:
\begin{equation}
\label{eq:CLsMultipleChannelsWoUncertAsympt}
\CLs=\frac{\Phi\left(\frac{\nobs_\text{eff}\left(\mu\right)-\sum\limits_c\beta_c\left(\mu\scnom+\bcnom\right)}{\sqrt{\sum\limits_c\beta_c^2\left(\mu\scnom+\bcnom\right)}}\right)}{\Phi\left(\frac{\nobs_\text{eff}\left(\mu\right)-\sum\limits_c\beta_c\bcnom}{\sqrt{\sum\limits_c\beta_c^2\bcnom}}\right)},
\end{equation}
where $\Phi$ is the cumulative distribution function of the standard normal distribution.
Eq. \ref{eq:muUpFromCLs} with Eq. \ref{eq:CLsMultipleChannelsWoUncertAsympt} can easily be solved by dichotomy.
This result has been used to validate the channel combination procedure implemented in \OTH~against the asymptotic limit.
Fig.~\ref{fig:ThreeChannelExampleAsympt} shows a comparison of this asymptotic result with \OTH~in the three channels example defined by
\begin{itemize}
\item channel 1: $\snom=5.18\times L$, $\bnom=2.22\times L$ and $\nobs=3\times L$
\item channel 2: $\snom=3.05\times L$, $\bnom=1.61\times L$ and $\nobs=4\times L$
\item channel 3: $\snom=4.45\times L$, $\bnom=2.95\times L$ and $\nobs=2\times L$
\end{itemize}

As can be seen from Fig.~\ref{fig:ThreeChannelExampleAsympt}, \OTH~converges, as expected, to the asymptotic result as the number of events increases.

\subsection{Validation of calculations with statistical and systematic uncertainties}
\label{sec:ValidationWithUncertainties}

\begin{figure*}[!htb]
\begin{center}
\includegraphics[scale=0.65]{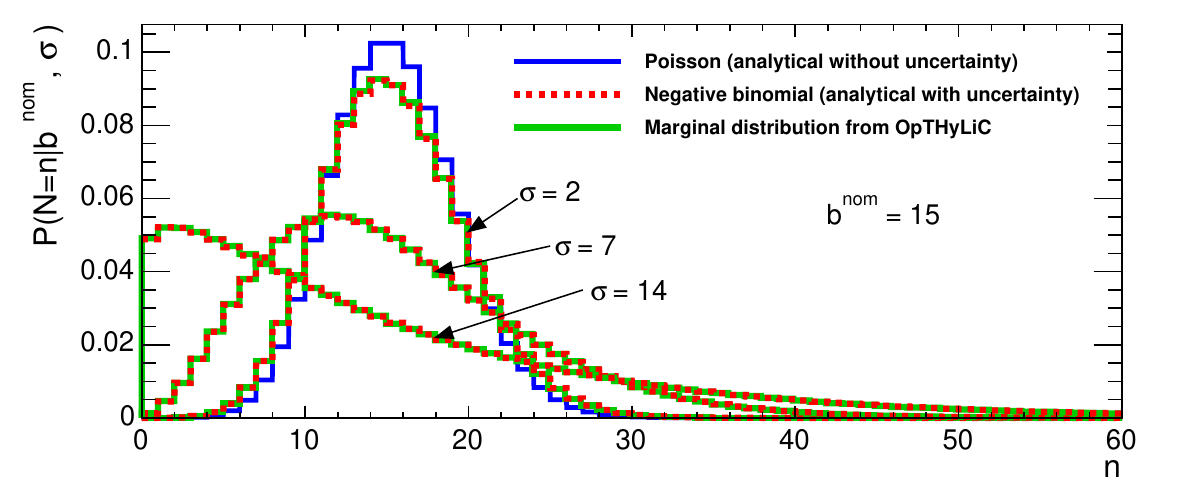}
\vspace{-2pt}
\caption{Marginal distribution of the yield under the background only hypothesis in the case where the background has a statistical uncertainty constrained by a gamma \pdf\ with mean $b^{\text{nom}}=15$ and three different values of standard deviation: $\sigma=2, 7$ and 14.\label{fig:SingleChannelStatUncertNegativeBinomial}}
\end{center}
\end{figure*}

Several other studies have been performed in order to validate the treatment of statistical and systematic uncertainties in \OTH.
As already explained, \OTH~treats uncertainties in a Bayesian way by marginalizing the likelihood.
A simple check of the marginalization procedure has been performed in the single channel case
by comparing marginal distributions of the yield as computed by \OTH~to analytical distributions
in the case where a single background affected by a statistical uncertainty constrained with a gamma \pdf\ contributes to the expected yield.
Indeed, it is known that, in this case, the marginal distribution of the yield,
given by the compound of a Poisson and a gamma distribution, is negative binomial:
\begin{equation}
\label{eq:NegativeBinomial}
 \begin{array}{rl}
P(N=n|b^{\text{nom}},\sigma)=&
\displaystyle\int_0^\infty P(N=n|b)\times f(b;b^{\text{nom}},\sigma)\dd b\\
&\\
=&\frac{\Gamma\left(N+\left(\frac{b^{\text{nom}}}{\sigma}\right)^2\right)}{N!\Gamma\left(\left(\frac{b^{\text{nom}}}{\sigma}\right)^2\right)}\left(\frac{b^{\text{nom}}}{b^{\text{nom}}+\sigma^2}\right)^{\left(b^{\text{nom}}/\sigma\right)^2}\hspace{-5pt}\left(\frac{\sigma^2}{b^{\text{nom}}+\sigma^2}\right)^{N},\\
 \end{array}
\end{equation}
where $N$ is the observed yield, $b$ is the background yield, $b^{\text{nom}}$ its nominal value,
$\sigma$ its statistical uncertainty, $P(N=n|b)$ the Poisson distribution with parameter $b$, $f(b;b^{\text{nom}},\sigma)$
the gamma distribution for $b$ with mean $b^{\text{nom}}$ and standard deviation $\sigma$
(given by Eq. \ref{eq:gammaPosteriorInv}) and $P(N=n|b^{\text{nom}},\sigma)$ the marginal (negative binomial) distribution of the yield $N$.
Fig. \ref{fig:SingleChannelStatUncertNegativeBinomial} shows that the agreement between Eq. \ref{eq:NegativeBinomial}
and marginal distributions calculated by \OTH~is very good.
Excellent agreement has also been observed for the two other gamma definitions available
in \OTH~(Eq.~\ref{eq:gammaPosteriorUni} and~\ref{eq:gammaPosteriorJeffrey}).

\begin{figure*}[!htb]
\begin{center}
\includegraphics[width=0.49\textwidth]{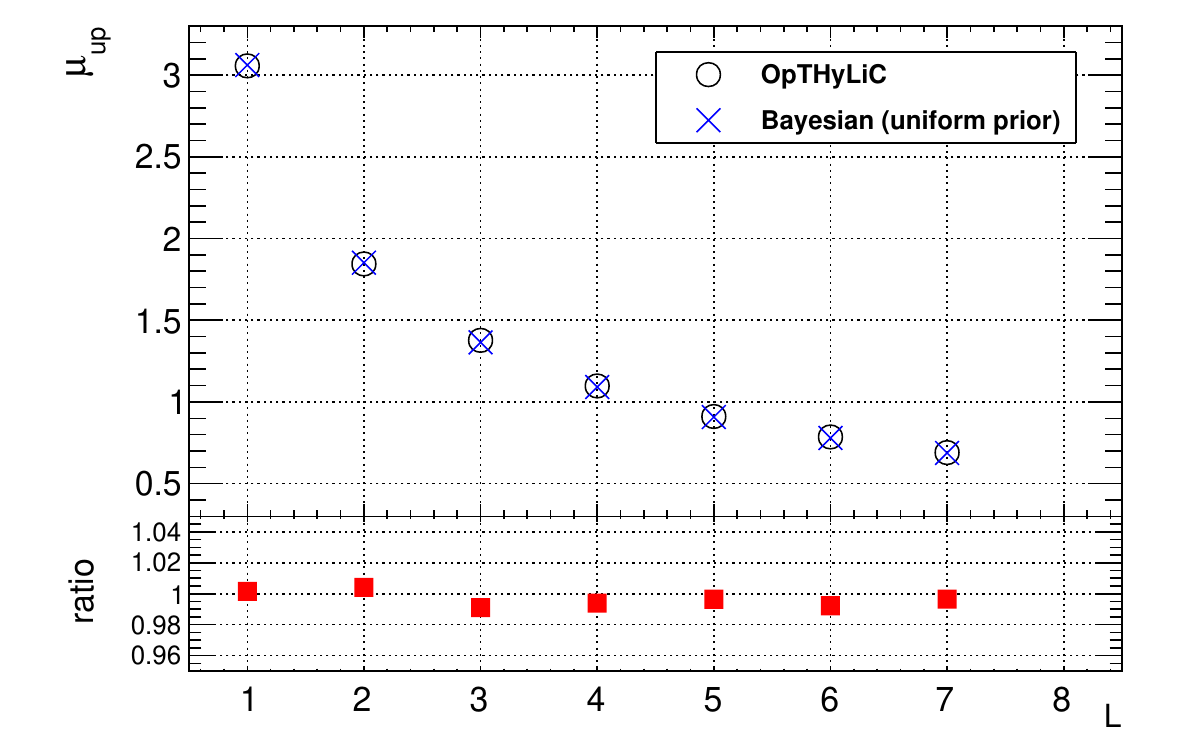}
\includegraphics[width=0.49\textwidth]{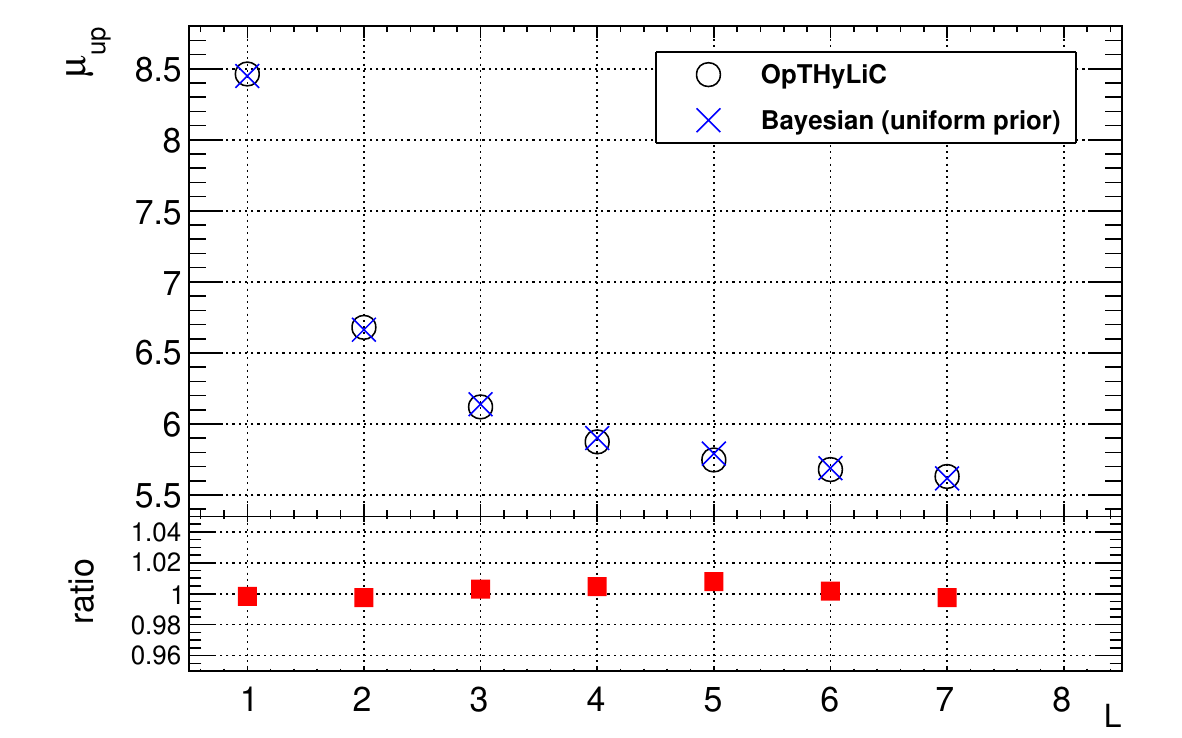}
\vspace{-2pt}
\caption{Upper limit \mup~as a function of $L$ without (left) and with (right) statistical and systematic uncertainties. It has been computed using exponential interpolation and extrapolation for systematic uncertainties and normal constraint terms for statistical uncertainty.\label{fig:ExampleValidUncert2}}
\end{center}
\vspace*{-20pt}
\end{figure*}

In order to validate not only the treatment of statistical uncertainties but also that of systematic uncertainties,
upper limits calculated with \OTH~have been compared to upper limits calculated using a Bayesian method.
It can indeed be shown that, in the single channel case, the hybrid \CLs{} method implemented
in \OTH~is equivalent to the Bayesian method when a uniform prior on the signal strength $\mu$
is used and when the signal has no uncertainties (neither statistical nor systematic) associated to it \cite{busato}.
This equivalence holds for any number of background sources, any number of statistical and systematic uncertainties associated to them
and any type of correlation between systematic uncertainties across background sources.
This result is used to complete the validation of the treatment of statistical uncertainties and to validate the treatment of systematic uncertainties.
In order to perform this validation, an independent Bayesian software, based on \roostats, has been developed.
This software implements exactly the same likelihood as \OTH~and takes as input the same files, hence allowing direct comparison to \OTH.
Several comparisons between \OTH~and Bayesian results have been performed
by changing the number of background sources and the number of systematic and statistical uncertainties.
One such comparison has been performed using the configuration given in Fig. \ref{fig:ExampleValidUncert1}, with $L=1,\hdots,7$.
The result is shown in Fig. \ref{fig:ExampleValidUncert2}.
\begin{figure}[!htb]
\centering\small
\begin{BVerbatim}
+bg Bkg1 25*L 7.
.syst Syst1 0.1 -0.3
.syst Syst2 0.3 -0.2

+bg Bkg2 25*L 12
.syst Syst1 0.2 -0.05
.syst Syst3 -0.06 0.15

+bg Bkg3 33.33*L 3.5
.syst Syst2 -0.1 0.3
.syst Syst3 0.15 -0.15
.syst Syst4 -0.6 0.6

+bg Bkg4 16.67*L 5
.syst Syst5 -0.3 0.25

+sig Sig 5*L 

+data 90*L
\end{BVerbatim}
\vspace{-2pt}
\caption{One of the examples used to validate the treatment of uncertainties in \OTH.\label{fig:ExampleValidUncert1}}
\end{figure}
This example has been chosen because uncertainties are large and their effect on upper limits is very pronounced
(as can be seen by comparing plots on the left and right of Fig. \ref{fig:ExampleValidUncert2}).
Thus, any mis-treatment of uncertainties in \OTH~should be visible.
Very good agreement between \OTH~and the Bayesian calculation with uniform prior is found for this example.
Similar agreements are found in other cases.
Rigorously, these comparisons validate only the treatment of statistical and systematic uncertaintes for backgrounds.
However, signal uncertainties are treated by the same code as background uncertainties.
Signal uncertainties are therefore expected to be treated properly.
As will be seen below, the comparison between \mclimit~and \OTH~gives confidence that signal uncertainties are indeed treated properly.
\clearpage

\begin{figure*}[!htb]
\begin{center}
\includegraphics[scale=0.6]{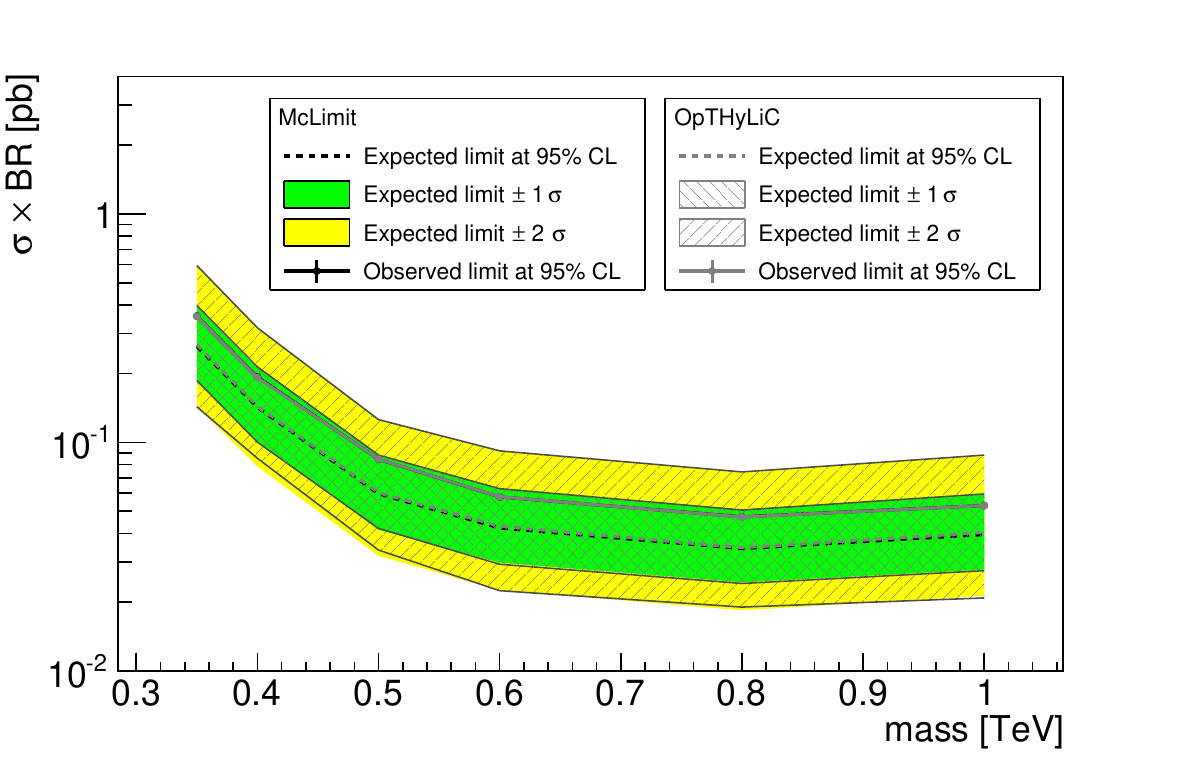}
\vspace{-2pt}
\caption{Comparison of observed and expected upper limits calculated with \mclimit~and \OTH~in a realistic case (see text). \label{fig:ExclusionPlot}}
\end{center}
\end{figure*}

The last validation compares observed and expected (\mbox{-2$\sigma$}, \mbox{-1$\sigma$}, \mbox{median}, \mbox{+1$\sigma$} and \mbox{+2$\sigma$})
upper limits calculated with \OTH~to upper limits calculated with \mclimit.
\mclimit, as \OTH, is a hybrid frequentist-Bayesian tool, using the interpolation/extrapolation described in Sec. \ref{sec:McLimitInterExtrap}
and normal constraints for statistical uncertainties.
When configured appropriately, \OTH~is therefore expected to give the same upper limits
as \mclimit\footnote{Profiling of uncertainties has been turned off in \mclimit~so as to allow direct comparison to \OTH.}.
For this comparison, typical inputs from high energy physics analysis have been used.
Fig. \ref{fig:ExclusionPlot} shows a comparison of upper limits
as a function of the mass of a hypothetical new particle obtained by combining three channels.
Six mass points have been considered and seven background processes contribute to the yield in each channel.
Signal and background processes are all affected by statistical and systematic uncertainties.
For each mass, the total number of nuisance parameters is 51 (27 are associated to systematic uncertainties and 24 to statistical uncertainties).
In both cases, 50 000 pseudo-experiments are used.
Limits found with \OTH~are in good agreement with those found with \mclimit.
The main difference between the two programs is the computing time.
For example, on the same computer, it took 25 minutes (8 seconds) to calculate the observed limit for the 1~TeV point with \mclimit~(\OTH).
The full plot is produced in less than 6 minutes with \OTH~while it takes several hours with \mclimit.

%% file: Conclusion.tex
\section{Conclusion}

A tool computing observed and expected limits has been presented.
This tool, named \OTH, is written in C++ and uses the \rootcern~library.
It implements the hybrid frequentist-Bayesian \CLs~method for hypothesis testing.
It can be used with an arbitrary number of channels and both for counting and multi-bin experiments.
Statistical and systematic uncertainties are accounted for as well as correlations between systematic uncertainties.
Several types of interpolation/extrapolation for systematic uncertainties and constraint terms for statistical uncertainties are provided.
\OTH~has been validated by comparing it to known analytical and Bayesian results and to \mclimit{} in situations where they are expected to give identical limits. Very good agreement has been found in all cases, hence validating the software.

One of the main advantages of \OTH{} is its speed.
Even in realistic cases with dozens of nuisance parameters and several channels and bins, the duration of limits computation remains of the order a few minutes or less.

%% file: Acknowledgements.tex
\section*{Acknowledgements}
The authors would like to thank R. Madar and L. Val\'ery for the fruitful discussions that helped improve the quality of this document.
They also would like to thank J. Linnemann for drawing their attention to the possibility of changing parameters of the log-normal distribution, extending the variety of priors that can be considered for statistical uncertainties.
Part of this work was supported by the regional council of Auvergne through the ``nouveau chercheur'' research funding program, and by the french ``Agence Nationale de la Recherche''.